\documentclass[aps,twocolumn,showpacs,preprintnumbers,amsmath,eqsecnum,floats,amssymb,nofootinbib]
{revtex4}
\usepackage{graphicx}
\usepackage{dcolumn}
\usepackage{bm}

\def\boldpi{\boldsymbol{\pi}}

\def\boldsigma{\boldsymbol{\sigma}}
\def\boldbeta{\boldsymbol{\beta}}
\def\boldmu{\boldsymbol{\mu}}
\def\boldxi{\boldsymbol{\xi}}

\def\sdotp{\boldsigma\cdot\boldpi}
\def\sdotE{\boldsigma\cdot\mathbf{E}}
\def\sdotB{\boldsigma\cdot\mathbf{B}}
\def\sep{\boldsigma\cdot(\mathbf{E}\times\boldsymbol{\pi})}
\def\mex{\boldmu\cdot(\mathbf{E}\times\boldxi)}
\def\mudotB{\boldmu\cdot\mathbf{B}}
\def\Bdotp{\mathbf{B}\cdot\boldpi}

\def\mathFW{\mathrm{FW}}
\def\Horbit{H_{\mathrm{orbit}}}
\def\Hspin{H_{\mathrm{spin}}}
\def\Hze{H_{\mathrm{ze}}}
\def\Hso{H_{\mathrm{so}}}

\def\boost{\widehat{\boldsymbol{\beta}}}
\def\gammaop{\widehat{\gamma}}

\newcommand{\secref}[1]{Sec.~\ref{#1}}

\begin{document}

\title{High-order Foldy-Wouthuysen transformations of the Dirac and Dirac-Pauli Hamiltonians in the weak-field limit}

\author{Tsung-Wei Chen}
\email{twchen@mail.nsysu.edu.tw}\affiliation{Department of
Physics, National Sun Yat-sen University, Kaohsiung 80424, Taiwan}

\author{Dah-Wei Chiou}
\email{dwchiou@gmail.com}
\affiliation{Department of Physics and Center for Condensed Matter Sciences, National Taiwan University, Taipei 10617, Taiwan}


\begin{abstract}
The low-energy and weak-field limit of Dirac equation can be obtained by an order-by-order block diagonalization approach to any desired order in the parameter $\boldpi/mc$ ($\boldpi$ is the kinetic momentum and $m$ is the mass of the particle). In the previous work, it has been shown that, up to the order of $(\boldpi/mc)^8$, the Dirac-Pauli Hamiltonian in the Foldy-Wouthuysen (FW) representation may be expressed as a closed form and consistent with the classical Hamiltonian, which is the sum of the classical relativistic Hamiltonian for orbital motion and the Thomas-Bargmann-Michel-Telegdi (T-BMT) Hamiltonian for spin precession. In order to investigate the exact validity of the correspondence between classical and Dirac-Pauli spinors, it is necessary to proceed to higher orders. In this paper, we investigate the FW representation of the Dirac and Dirac-Pauli Hamiltonians by using Kutzelnigg's diagonalization method. We show that the Kutzelnigg diagonalization method can be further simplified if nonlinear effects of static and homogeneous electromagnetic fields are neglected (in the weak-field limit). Up to the order of $(\boldpi/mc)^{14}$, we find that the FW transformation for both Dirac and Dirac-Pauli Hamiltonians is in agreement with the classical Hamiltonian with the gyromagnetic ratio given by $g=2$ and $g\neq2$ respectively. Furthermore, with higher-order terms at hand, it is demonstrated that the unitary FW transformation admits a closed form in the low-energy and weak-field limit.
\end{abstract}


\maketitle

\section{Introduction}

The relativistic quantum theory for a spin-1/2 particle is
described by a spinor satisfying the Dirac equation
\cite{Dirac1928, Dirac1982}. The four-component spinor of the
Dirac particle is composed of two two-component Weyl spinors which
correspond to the particle and antiparticle parts. Rigourously,
because of the non-negligible probability of creation/annihilation
of particle-antiparticle pairs, the Dirac equation is
self-consistent only in the context of quantum field theory. For
the purpose of obtaining the low-energy limit of Dirac equation
without accounting for the field-theory particle-antiparticle
interaction, the Dirac equation is converted to a two-component
equation.

The Pauli substraction method eliminates the two small components
from the four-component spinor of Dirac equation and leads to the
block-diagonal but energy-dependent effective Hamiltonian in which
some non-hermitian terms may appear. Apart from the difficulties,
in the seminal paper \cite{Foldy1950}, Foldy and Wouthuysen (FW)
established a series of successive unitary transformations via
decomposing the Hamiltonian into even and odd matrices; a
block-diagonalized effective Hamiltonian can be constructed up to
a certain order of $\boldpi/mc$. The series of successive unitary
transformations in the FW method can be replaced by a single
transformation via the L\"{o}wding partitioning method
\cite{Lowdin1951,Winkler2003}.\footnote{It should be emphasized that The FW
transformation is not meant to be used for the second
quantization, and furthermore it exists only in the weak-field
limit or in some special cases (as studied in \secref{sec:exact solution}). If we
ruthlessly try to second quantize the theory in the FW
representation, whether we can succeed or not, we should perform the second quantization in a way
very different from the conventional approach. This is because, in
the FW representation, we encounter the non-locality due to
\emph{zitterbewegung} (see also P.\ Strange in
Ref.~\cite{Winkler2003}).}

Furthermore, Eriksen developed a systematic derivation of the
unitary transformation and gave an exact FW transformation for a
charged spin-1/2 particle in interaction with non-explicitly
time-dependent field~\cite{Eriksen1958}. The validity of the
Eriksen method is investigated in Ref.\ \cite{Vries1968}. In Ref.\
\cite{Kutz1990}, Kutzelnigg developed a single unitary
transformation that allows one to obtain the block-diagonalized
Dirac Hamiltonian without evoking the decomposition of even and
odd matrices used in FW method. Alternatively, the Dirac
Hamiltonian can also be diagonalized via expansion in powers of
Planck constant $\hbar$ \cite{Silenko03,Bliokh05,Goss07}, which
enables us to investigate the influences of quantum corrections on
the classical dynamics in strong fields \cite{Silenko08}.

On the other hand, the classical relativistic dynamics for a
charged particle with intrinsic spin in static and homogeneous
electromagnetic fields is well understood. The orbital motion is
governed by the classical relativistic Hamiltonian
\begin{equation}\label{H orbit}
H^c_\mathrm{orbit} =\sqrt{c^2\boldpi^2+m^2c^4}\, +V(\mathbf{x}),
\end{equation}
where $\boldpi=\mathbf{p}-q\mathbf{A}/c$ is the kinetic momentum
operator with $\mathbf{A}$ being the magnetic vector
potential~\cite{Jackson} and $V(\mathbf{x})$ the electric
potential energy. The spin motion is governed by the
Thomas-Bargmann-Michel-Teledgi (T-BMT) equation which describes
the precession of spin as measured by the laboratory
observer~\cite{BMT59},
\begin{equation}\label{Thomas}
\frac{d\mathbf{s}}{dt}=\frac{q}{mc}\,\mathbf{s}\times\mathbf{F}(\mathbf{x})
\end{equation}
with
\begin{equation}\label{ThomasF}
\begin{split}
\mathbf{F}&=\left(\frac{g}{2}-1+\frac{1}{\gamma}\right)\mathbf{B}-\left(\frac{g}{2}-1\right)\frac{\gamma}{\gamma+1}(\boldsymbol{\beta}\cdot\mathbf{B})\boldsymbol{\beta}\\
&~~-\left(\frac{g}{2}-\frac{\gamma}{\gamma+1}\right)\boldsymbol{\beta}\times\mathbf{E},
\end{split}
\end{equation}
where $g$ is the gyromagnetic ratio, $\boldbeta$ the boost
velocity, $\gamma=1/\sqrt{1-\boldbeta^2}$ the Lorentz factor and
$\mathbf{E}$ and $\mathbf{B}$ are electric and magnetic fields
measured in the laboratory frame. The intrinsic spin $\mathbf{s}$
in Eq.~(\ref{Thomas}) is being observed in the rest frame of the
particle. Because $\{s_i,s_j\}=\epsilon_{ijk}s_k$,
Eq.~(\ref{Thomas}) can be recast as the Hamilton's equation:
\begin{equation}
\frac{d\mathbf{s}}{dt}=\{\mathbf{s},H^{\mathrm{c}}_\mathrm{spin}\}
\end{equation}
with
\begin{equation}\label{H dipole}
H^{\mathrm{c}}_\mathrm{spin}=-\frac{q}{mc}\,\mathbf{s}\cdot
\mathbf{F}
\end{equation}
called the T-BMT Hamiltonian. The combination of Eqs (\ref{H
orbit}) and (\ref{H dipole}) is hereafter called the classical
Hamiltonian $H_{\mathrm{c}}$,
\begin{equation}\label{H classical}
\begin{split}
H_{\mathrm{c}}&=H^{\mathrm{c}}_\mathrm{orbit}+H^{\mathrm{c}}_\mathrm{spin}\\
&=\sqrt{c^2\boldpi^2+m^2c^4}\, +V-\boldmu\cdot \mathbf{F},
\end{split}
\end{equation}
where $\boldmu=q\mathbf{s}/mc$ is the intrinsic magnetic moment of
an electron.

The connection between the Dirac equation and classical
Hamiltonian has been investigated by several authors
\cite{Foldy1950, Rubinow1963, Rafa1964, Froh1993, Silenko1995}.
For a free Dirac particle, it has been shown that the exactly
diagonalized Dirac Hamiltonian corresponds to the classical
relativistic Hamiltonian \cite{Foldy1950}. In
Refs.~\cite{Rubinow1963,Rafa1964}, it was shown that the T-BMT
equation may be derived from the WKB wavefunction solutions to the
Dirac equation. In the presence of external electromagnetic
fields, the Dirac Hamiltonian in the FW representation has been
block-diagonalized up to the order of $(\boldpi/mc)^4$, but the
connection is not explicit \cite{Froh1993}. Recently, in
Ref.~\cite{TWChen2010}, it has been shown that up to
$(\boldpi/mc)^8$, the resulting FW transformed Dirac, or more
generic, Dirac-Pauli \cite{Pauli1941} Hamiltonian in the presence
of static and homogeneous electromagnetic fields
may agree with the classical
Hamiltonian [Eq.~(\ref{H classical})] in the weak-field limit.


The order-by-order block-diagonalization methods to higher orders of
$\boldpi/mc$ can be used to investigate the validity of the
connection. Furthermore, if the connection is indeed
establishable (in a closed form), corrections to the classical T-BMT equation due to
field inhomogeneity, if any, could also be included. Motivated by these
regards, we adopt a systematic method that can substantially simplify the
calculation of FW transformation to any higher
orders in the FW representation of Dirac Hamiltonian. It must be
stressed that block diagonaliation of a four-component Hamiltonian
into two uncoupled two-component Hamiltonian is not unique, as any composition with additional
unitary transformations that act separately on the positive and negative energy blocks will also do the job. Different block-diagonalization transformations are however unitarily equivalent to one another, and thus yield the same physics.\footnote{Once the Hamiltonian is block-diagonalized, further unitary transformations that do not mix the positive and negative energy blocks merely rotate the $2\times2$ Pauli matrices independently for the two blocks, keeping the physics unchanged.} The truly vexed question is: whether does the unitary bock-diagonalization transformation exist at all? In the absence of electric fields, we will show that the answer is affirmative. On the other hand, in the presence of electric fields, the answer seems to be negative, as the energy interacting with electromagnetic fields renders the probability of creation/annihilation of particle-antiparticle pairs non-negligible and thus the particle-antiparticle separation inconsistent. Nevertheless, in the weak-field limit, the interacting energy is well below the Dirac energy gap ($2mc^2$) and we will demonstrate that the unitary transformation exists and indeed admits a closed form in the low-energy and weak-field limit.

In this article, we derive the FW transformed Dirac Hamiltonian up
to the order of $(\boldpi/mc)^{14}$ by using Kutzelnigg's
diagonalization method \cite{Kutz1990}. The key feature of the
Kutzelnigg approach is that it provides an exact
block-diagonalized form of Dirac Hamiltonian involving a
self-consistent equation [see Eq.~(\ref{EqX})]. The explicit form
of the FW transformed Dirac Hamiltonian can be obtained by solving
the self-consistent equation. We will show that the Kutzelnigg
method can be further simplified in the weak-field
limit, and this simplification enables us to obtain the
higher-order terms systematically. We will show that the
block diagonalization of Dirac and Dirac-Pauli Hamiltonians
up to the order of $(\boldpi/mc)^{14}$ in the Foldy-Wouthuysen
representation is in agreement with classical Hamiltonian, and the
closed form of the unitary transformation can be found.


This article is organized as follows. In Sec.~\ref{sec:method}, we
construct a unitary operator based on the Kutzelnigg method to
obtain the exact FW transformed Dirac Hamiltonian and the
self-consistent equation. The exact solution of the
self-consistent equation is discussed. The FW transformed Dirac
Hamiltonian in the presence of inhomogeneous electromagnetic
fields are derived in Sec.~\ref{sec:fields}. The effective
Hamiltonian up to $(\boldpi/mc)^4$ for the inhomogeneous
electromagnetic field is in agreement with the previous result
shown in Refs.~\cite{Foldy1950,Froh1993}. The static and
homogeneous electromagnetic fields are considered in
Sec.~\ref{sec:HFW}, where the simplification of the effective
Hamiltonian is discussed and the FW transformed Dirac Hamiltonian
is obtained up to $(\boldpi/mc)^{14}$. In Sec.~\ref{sec:TBMT}, the
comparison with the classical relativistic Hamiltonian and T-BMT
equation with $g=2$ is discussed. The FW transformed Dirac-Pauli
Hamiltonian is shown in Sec.~\ref{sec:TBMT2}. In
Sec.~\ref{sec:EUT}, we demonstrate that the exact unitary
transformation matrix in the low-energy and weak-field limit
can be formally obtained. The conclusions are summarized in
Sec.~\ref{sec:conclusions}. Some calculational details are
supplemented in Appendices.

\section{Kutzelnigg diagonalization method for Dirac Hamiltonian}\label{sec:method}
In this section, we use the unitary operator based on the
Kutzelnigg diagonalization method \cite{Kutz1990} and apply the
unitary operator to the Dirac Hamiltonian. We obtain the
formally exact Foldy-Wouthuysen transformed Hamiltonian by
requiring that the unitary transformation yields a block-diagonal
form.


The Dirac Hamiltonian in the presence of electromagnetic fields
can be written as
\begin{equation}\label{Dirac}
\begin{split}
H_D&=\left(\begin{array}{cc} V+mc^2& c\boldsigma\cdot\boldpi\\
c\boldsigma\cdot\boldpi&V-mc^2
\end{array}\right)\\
&\equiv\left(\begin{array}{cc}h_+& h_0\\
h_0&h_-
\end{array}\right)
\end{split}
\end{equation}
where $\boldpi=\mathbf{p}-q\mathbf{A}/c$ is the kinetic momentum
operator and $V=q\phi$. The electric field and magnetic field are
$\mathbf{E}=-\nabla\phi$ and $\mathbf{B}=\nabla\times\mathbf{A}$,
respectively. We note that in the static case,
$\nabla\times\mathbf{E}=0$, and thus,
$\boldpi\times\mathbf{E}=-\mathbf{E}\times\boldpi$. The wave
function of the Dirac equation
$H_D\psi=i\hbar\frac{\partial}{\partial t}\psi$ is a two
two-spinors
\begin{equation}
\psi=\left(\begin{array}{c} \psi_+\\
\psi_-
\end{array}\right).
\end{equation}
A unitary operator $U$ which \emph{formally} decouples positive and negative
energy states can be written as the following form \cite{Kutz1990}
\begin{equation}\label{U}
U=\left(\begin{array}{cc} Y&YX^{\dag}\\
-ZX&Z
\end{array}\right),
\end{equation}
where operators $Y$ and $Z$ are defined as
\begin{equation}\label{Def:YandZ}
Y=\frac{1}{\sqrt{1+X^{\dag}X}}, \qquad
Z=\frac{1}{\sqrt{1+XX^{\dag}}}.
\end{equation}
Applying the unitary transformation Eq.~(\ref{U}) to
Eq.~(\ref{Dirac}), $UH_DU^{\dag}$ is of the form
\begin{equation}\label{UHU}
UH_DU^{\dag}=\left(\begin{array}{cc} H_{\mathFW}&H_{X^{\dag}}\\
H_{X}&H'
\end{array}\right)
\end{equation}
The unitary transformation transforms the wave function
$\psi$ to a two-spinor,
\begin{equation}
U\left(\begin{array}{c} \psi_+\\
\psi_-
\end{array}\right)=\left(\begin{array}{c} \psi_{\mathrm{FW}}\\
0
\end{array}\right),
\end{equation}
where the wave function for the negative energy state must be zero
and the FW transformed wave function is given by
\begin{equation}
\psi_{\mathrm{FW}}=\sqrt{1+X^{\dag}X}\,\psi_+.
\end{equation}
We require that the transformed Hamiltonian takes the
block-diagonal form:
\begin{equation}\label{UHU-D}
UH_DU^{\dag}=\left(\begin{array}{cc} H_{\mathFW}&0\\
0&H'
\end{array}\right).
\end{equation}
We find that the requirement of the vanishing off-diagonal term
$H_{X}=0$ yields the constraint on the $X$ operator:
\begin{equation}\label{EqX}
X=\frac{1}{2mc^2}\left\{-Xh_0X+h_0+[V,X]\right\}.
\end{equation}
Equation (\ref{EqX}) is a self-consistent formula for operator
$X$. The resulting FW transformed Hamiltonian
$H_{\mathrm{FW}}$ is given by
\begin{equation}\label{Un-HFW}
\begin{split}
H_{\mathFW}=Y\left(h_++X^{\dag}h_0+h_0X+X^{\dag}h_-X\right)Y.
\end{split}
\end{equation}
Because the operator $X$ plays an important role in generating the
FW transformed Hamiltonian and the corresponding unitary operator,
the operator $X$ for the Dirac Hamiltonian is hereafter called the
\emph{Dirac generating operator}. To our knowledge, the exact
solution of Eq.\ (\ref{EqX}) for a general potential is
still unknown except for the two cases: a free particle and a
particle subject only to magnetic fields. For the case with a nontrivial
electric potential, we assume that the solution of Eq. (\ref{EqX})
can be obtained by using series expansion.\footnote{The series solution of the Dirac generating operator $X$ is not unique because any
unitary transformation would lead to a satisfactory $X$ as long as it does not mix positive and negative
energy states. This implies that the form of the block-diagonalized Hamiltonian is
not unique. In this regard, we focus only on the series solution of
$X$ that can correctly generate the FW transformed Dirac
Hamiltonian linear in EM fields and up to order of $(\boldpi/mc)^4$,
as shown in Sec.\ref{sec:fields}.}

\subsection{Exact solution of Dirac generating operator}\label{sec:exact solution}
For a free particle ($\mathbf{A}=0$ and $V=0$), it can be
shown that Eq. (\ref{EqX}) has an exact solution
\begin{equation}\label{free-X}
X=\frac{c(\boldsigma\cdot\mathbf{p})}{mc^2+E_p},
\end{equation}
where $E_p=\sqrt{m^2c^4+\mathbf{p}^2c^2}$. Using Eqs. (\ref{U})
and (\ref{Def:YandZ}), the unitary transformation matrix can be
written as
\begin{equation}\label{U-free}
U=\frac{1}{\sqrt{2E_p(E_p+mc^2)}}\left(\begin{array}{cc}
E_p+mc^2&c\boldsigma\cdot\mathbf{p}\\
-c\boldsigma\cdot\mathbf{p}&E_p+mc^2\\
\end{array}\right).
\end{equation}
Equation (\ref{U-free}) is the same with the result obtained from
the standard FW transformation~\cite{Foldy1950}.
The resulting FW transformed free-particle Dirac Hamiltonian is block-diagonalized,
\begin{equation}\label{FWfree}
H_{\mathrm{FW}}=\left(\begin{array}{cc}
E_p&0\\
0&-E_p\\
\end{array}\right).
\end{equation}
It is interesting to note that in the absence of electric field
(i.e., $V=\text{const}$), Eq. (\ref{EqX}) also
admits an exact solution
\begin{equation}\label{Magnetic-X}
X=\frac{1}{mc^2+E_{\pi}}(c\boldsigma\cdot\boldpi),
\end{equation}
where $E_{\pi}=\sqrt{m^2c^4+c^2(\boldsigma\cdot\boldpi)^2}$. This
can be proved by directly substituting Eq. (\ref{Magnetic-X}) into
Eq. (\ref{EqX}) with $V=\text{const}$. The exact unitary transformation
matrix can be formally constructed and the resulting FW transformed
Hamiltonian can be obtained. In the presence of a nontrivial electric
potential, it is difficult to obtain an exact solution because the
term $[V,X]$ does not vanish. Therefore, the diagonalization
procedure for the Dirac Hamiltonian must be performed order-by-order.
It is necessary to choose a dimensionless quantity as the
order-expanding parameter. We note that the form of Eq.
(\ref{Magnetic-X}) can be rewritten as
$\{1/[1+(\boldsigma\cdot\boldxi)^2]\}\boldsigma\cdot\boldxi$ with
the dimensionless quantity $\boldxi=\boldpi/mc$. The
order-by-order block-diagonalized Hamiltonian can be expressed in
terms of the order parameter $\boldxi$. In this paper, we
further focus on the weak-field limit in order to compare
the block-diagonalizaed Hamiltonian with the classical counterpart.

\subsection{Series expansion of Dirac generating operator}

The upper-left diagonal term of $UH_DU^{\dag}$ is the FW
transformed Hamiltonian $H_{\mathFW}$ under the constraint
Eq.~(\ref{EqX}) and it can be written as (see Appendix
\ref{App:Ham})
\begin{equation}\label{HFW}
\begin{split}
H_{\mathFW}=mc^2+e^{G/2}Ae^{-G/2},
\end{split}
\end{equation}
where operators $A$ (hereafter called the \emph{Dirac energy
operator}) and $G$ (hereafter called the \emph{Dirac exponent
operator}) are defined as
\begin{equation}\label{AG}
A\equiv V+h_0X,
\qquad
G\equiv\ln\left(1+X^{\dag}X\right).
\end{equation}
The other requirement $H_{X^{\dag}}=0$ gives the constraint on the
hermitian of the Dirac generating operator $X^{\dag}$, which is
simply the hermitian conjugate of Eq.~(\ref{EqX}), namely,
$H_{X}^{\dag}=H_{X^{\dag}}$. It can be shown that $H_{X}=0$ and
$H_{X^{\dag}}=0$ imply that (see Appendix \ref{App:Ham})
\begin{equation}\label{HFWdag}
H_{\mathFW}^{\dag}=mc^2+e^{-G/2}A^{\dag}e^{G/2}
\end{equation}
and then Eq.~(\ref{HFW}) is a hermitian operator since the Dirac
exponent operator is a hermitian operator.  Equation (\ref{HFW})
can be simplified if we rewrite the Dirac energy operator $A$ as
the sum of its hermitian ($A^H$) and anti-hermitian ($A^N$) parts,
$A=A^H+A^N$, where
\begin{equation}
\begin{split}
A^H=\frac{A+A^{\dag}}{2},
\qquad
A^N=\frac{A-A^{\dag}}{2}.
\end{split}
\end{equation}
Combining Eqs.~(\ref{HFW}) together with (\ref{HFWdag}), the FW
transformed Dirac Hamiltonian $H_{\mathFW}$ is made up of
$H_{\mathFW}=\left(H_{\mathFW}+H_{\mathFW}^{\dag}\right)/2$, and
it can be written as
\begin{equation}\label{HFW NE}
H_{\mathFW}=mc^2+A^H+S,
\end{equation}
where the \emph{Dirac string operator} $S$ is given by
\begin{equation}\label{EqS NE}
\begin{split}
S&=\frac{1}{2}[G,A^N]+\frac{1}{2!2^2}[G,[G,A^H]]+\frac{1}{3!2^3}[G,[G,[G,A^N]]]\\
&~~+\frac{1}{4!2^4}[G,[G,[G,[G,A^H]]]]+\cdots,
\end{split}
\end{equation}
where we have used the Baker-Campbell-Hausdorff formula
\cite{Sak}:
$e^{B}De^{-B}=D+[B,D]+[B,[B,D]]/2!+[B,[B,[B,D]]]/3!\cdots$. We
note that the anti-hermitian part of the Dirac energy operator
always appears in those terms with odd numbers of Dirac exponent
operators, and the hermitian part of the Dirac energy operator
always appears in those terms with even numbers of Dirac exponent
operators. Since the commutator of two hermitian operators must be
an anti-hermitian operator, it can be shown that the Dirac string
operator is a hermitian operator.

In order to compare FW transformed Dirac Hamiltonian to the
classical Hamiltonian, we solve the self-consistent equation for
$X$ [Eq.~(\ref{EqX})] by power series expansion in terms of orders
of $1/c$,
\begin{equation}\label{SeriesX}
X=\frac{X_1}{c}+\frac{X_2}{c^2}+\frac{X_3}{c^3}+\cdots.
\end{equation}
$X_1$ is the first order of the Dirac generating operator, $X_2$
the second, and so on. Substituting Eq.~(\ref{SeriesX}) into
Eq.~(\ref{EqX}), we can obtain each order of the Dirac generating
operator $X_{\ell}$. For order of $1/c$ and $1/c^2$, we have
\begin{equation}\label{App:EqX1}
\begin{split}
&2mX_1=\sdotp,\\
&2mX_{2}=0.
\end{split}
\end{equation}
The expanding terms of the Dirac generating operator with third
and higher orders can be determined by the following equations,
\begin{equation}\label{App:EqXoe}
\begin{split}
2mX_{2j}&=-\sum_{k_1+k_2=2j-1}X_{k_1}\sdotp X_{k_2}+[V,X_{2j-2}],\\
2mX_{2j+1}&=-\sum_{k_1+k_2=2j}X_{k_1}\sdotp X_{k_2}+[V,X_{2j-1}],
\end{split}
\end{equation}
where $j=1,2,3\cdots$. Consider terms of even order of $1/c$,
namely, $X_{2j}$. The fourth order of the Dirac generating
operator $X_{4}$ is determined by $2mX_4=-(X_{1}\sdotp
X_{2}+X_{1}\sdotp X_{2})+[V,X_2]$. Since the second order of $X$
is zero ($X_2=0$), the fourth order of $X$ also vanishes, i.e.,
$X_4=0$. The sixth order of $X$ is obtained by
$2mX_6=-(X_{1}\sdotp X_{4}+X_{2}\sdotp X_{3}+X_{3}\sdotp
X_{2}+X_{4}\sdotp X_{1})+[V,X_4]$. Because both the second and
fourth order of the Dirac generating operators vanish, we find
that the sixth order of Dirac generating operator $X_6$ is also
zero, as well as $X_8$, $X_{10}$, and so on. Therefore, we have
\begin{equation}
X_2=X_4=X_6=\cdots=0,
\end{equation}
and the non-zero terms are those expanding terms of the Dirac
generating operators with odd subscripts, namely,
$X=X_1/c+X_3/c^3+X_5/c^5+\cdots$. Furthermore, since the operator
$h_0=c\sdotp$ is of the order of $c$, the series expansion of
$A=V+h_0X$ has only even powers of $c$:
\begin{equation}
A=A_0+\frac{A_2}{c^2}+\frac{A_4}{c^4}+\cdots,
\end{equation}
where the $\ell$th order of the Dirac energy operator $A_{\ell}$
is related to the $\ell$th order of Dirac generating operator
$X_{\ell}$ by
\begin{equation}\label{EqA}
\begin{split}
&A_0=V+\frac{h_0}{c}X_1,\\
&A_{\ell}=\frac{h_0}{c}X_{\ell+1},
\end{split}
\end{equation}
where $\ell=2,4,6,\cdots$. On the other hand, the series expansion
of $\ln(1+y)$ is $\ln(1+y)=y-y^2/2+y^3/3-y^4/4+\cdots$. Because
$y=X^{\dag}X$, the power series of $y$ contains only even powers
of $c$: $y=y_2/c^2+y_4/c^4+y_6/c^6+\cdots$, where $y_{\ell}$ are
given by
\begin{equation}\label{y ell}
y_{\ell}=\sum_{k_1+k_2=\ell}X^{\dag}_{k_1}X_{k_2}.
\end{equation}
For example, $y_6=X_1^{\dag}X_5+X_3^{\dag}X_3+X_5^{\dag}X_1$.
Consequently, the Dirac exponent operator can only have terms with
even powers of $c$ (we note that $G_0=0$)
\begin{equation}
G=\frac{G_2}{c^2}+\frac{G_4}{c^4}+\frac{G_6}{c^6}+\cdots,
\end{equation}
where the $\ell$th order of the Dirac exponent operator $G_{\ell}$
can be expressed in terms of $y_{\ell}$:
\begin{equation}\label{G ell}
\begin{split}
G_{\ell}&=y_{\ell}-\frac{1}{2}\sum_{k_1+k_2=\ell}y_{k_1}y_{k_2}+\frac{1}{3}\sum_{k_1+k_2+k_3=\ell}y_{k_1}y_{k_2}y_{k_3}\\
&~~-\frac{1}{4}\sum_{k_1+\cdots+k_4=\ell}y_{k_1}y_{k_2}y_{k_3}y_{k_4}+\cdots.
\end{split}
\end{equation}
For example, $G_6=y_6-(y_2y_4+y_4y_2)/2+y_2^3/3$. Therefore, the
FW transformed Dirac Hamiltonian can be expanded in terms of
$A_{\ell}$ and $G_{\ell}$ and has only even powers of $c$. That
is,
\begin{equation}\label{HFW sum}
H_{\mathFW}=mc^2+\sum_{\ell}H^{(\ell)}_{\mathFW},
\end{equation}
where the $\ell$th order of the FW transformed Dirac Hamiltonian
denoted as $H_{\mathFW}^{(\ell)}$ ($\ell=0,2,4,6,\cdots$) are
given by (up to $c^{12}$)
\begin{equation}\label{EqExpandH}
\begin{split}
H_{\mathFW}^{(0)}&=A^H_0,\\
c^{\ell}H_{\mathFW}^{(\ell)}&=A^H_{\ell}+S_{\ell},\\
\end{split}
\end{equation}
where $\ell=2,4,6,\cdots,12$. The $\ell$th order of Dirac string
operator $S_{\ell}$ is given by
\begin{equation}\label{EqS}
\begin{split}
&S_{\ell}=\frac{1}{2}\mathop{\sum_{\ell_1+\ell_2=\ell}}[G_{\ell_1},A^N_{\ell_2}]\\
&~~+\frac{1}{2!2^2}\sum_{\ell_1+\ell_2+\ell_3=\ell}[G_{\ell_1},[G_{\ell_2},A^H_{\ell_3}]]\\
&~~+\frac{1}{3!2^3}\sum_{\ell_1+\cdots+\ell_4=\ell}[G_{\ell_1},[G_{\ell_2},[G_{\ell_3},A^N_{\ell_4}]]]\\
&~~+\frac{1}{4!2^4}\sum_{\ell_1+\cdots+\ell_5=\ell}[G_{\ell_1},[G_{\ell_2},[G_{\ell_3},[G_{\ell_4},A^H_{\ell_5}]]]]\\
&~~+\cdots.
\end{split}
\end{equation}

As mentioned above, any unitary transformation would lead to
a satisfactory generating operator as long as it does not mix
positive and negative energy states. The non-uniqueness property
of generating operator can be easily seen as follows. If we
perform the Kutzelnigg diagonalization method upon Eq.\ (\ref{UHU-D})
again, then we obtain another block-diaogonalized Hamiltonian with
new operator equation for the generating operator. The new
diagonalized Hamiltonian $H_{\mathrm{FW}}'$ is determined by Eq.\
(\ref{Un-HFW}) with the replacements: $h_-\rightarrow H'$,
$h_0=0$, and $h_+ \rightarrow H_{\mathrm{FW}}$. The form of new
diagonalized Hamiltonian depends on the solution of the new
generating operator.

We use series expansion to construct the generating operator
and require that the resulting generating operator can go back to
the exact solution in the free-particle case where the Hamiltonian
is block-diagonalized to Eq. (\ref{FWfree}). In this
representation, the positive and negative energies are decoupled
and have classical relativistic energy representation [\textit{c.f.}\ Eqs.\
(\ref{FWfree}) and (\ref{H orbit})], which is \emph{the} FW
representation obtained in this article. Importantly, we will show that the
series expansion of generating operator [Eq.\ (\ref{SeriesX})] can
indeed generate the FW representation. In this sense,
interestingly, we can obtain an exact solution of generating
operator and find that the spin part of the resulting
block-diagonalized Hamiltonian is equivalent to the T-BMT
Hamiltonian.

In the next section, we will show that by using Eqs.\
(\ref{EqExpandH}) and (\ref{EqS}) the effective Hamiltonian
resulting from Foldy-Wouthuysen diagonalization method is
equivalent to that from the Kutzelnigg diagonalization method up
to terms with order of $(\boldpi/mc)^4$, from which the fine
structure, Darwin term and spin-orbit interaction can be deduced.

\section{Inhomogeneous fields}\label{sec:fields}
Up to this step, only two assumptions are made: (1) the
electromagnetic fields are static, and (2) the Dirac generating
operator $X$ can be solved by series expansion. We calculate the
first two terms $H^{(0)}_{\mathFW}+H^{(2)}_{\mathFW}$ and show
that the resulting Hamiltonian
$H_{\mathFW}=mc^2+H^{(0)}_{FW}+H^{(2)}_{\mathFW}$ is in agreement
with the previous result. The zeroth order of the FW transformed
Dirac Hamiltonian is
\begin{equation}\label{H0FW_1}
H_{\mathFW}^{(0)}=A^H_0,
\end{equation}
where $A_0=V+(h_0/c)X_1$. The first order of the Dirac generating
operator $X_1$ is given in Eq.~(\ref{App:EqX1}), which is valid
for inhomogeneous fields. Using
$[\pi_i,\pi_j]=\frac{iq\hbar}{c}\epsilon_{ijk}B_k$, we have
$(\sdotp)^2=\boldpi^2-\frac{q\hbar}{c}\sdotB$, and $H_{FW}^{(0)}$
[Eq.~(\ref{H0FW_1})] becomes
\begin{equation}\label{H0FW_2}
H^{(0)}_{\mathFW}=V+\frac{\boldpi^2}{2m}-\frac{q\hbar}{2mc}\sdotB.
\end{equation}
We note that $A_0$ is already a hermitian operator, and thus
$A^N_0=0$. The second and third terms of Eq.~(\ref{H0FW_2}) are
the kinetic energy and Zeeman energy. The second order
$H_{\mathFW}^{(2)}$ is given by
\begin{equation}\label{H2FW_1}
c^2H^{(2)}_{\mathFW}=A^H_2+S_2,
\end{equation}
where $A_2=(h_0/c)X_3$, $S_2=[G_2,A_0^N]/2$. For $X_3$, from Eq.~(\ref{App:EqXoe}) we have $X_3=-X_1\sdotp X_1+[V,X_1]$. Using
$[V,\,\sdotp]=-iq\hbar\sdotE$, we obtain
\begin{equation}\label{App:EqX3}
X_3=-\frac{1}{4}\frac{T}{m^2}\sdotp-\frac{1}{4}\frac{i\hbar}{m^2}\boldsigma\cdot\mathbf{E},
\end{equation}
where
\begin{equation}
T\equiv(\sdotp)^2/2m
\end{equation}
is the kinetic energy operator. The operators $X_3$ is valid for
inhomogeneous fields. From Eqs.~(\ref{y ell}) and (\ref{G ell}),
the operator $G_2$ is $X^{\dag}_1X_1=T/2m$. Since $A^N_0=0$, we
have $S_2=[G_2,A^N_0]/2=0$. Substituting Eq.~(\ref{App:EqX3}) into
$A_2$, we have
\begin{equation}
A_2=-\frac{(\sdotp)^4}{8m^3}-\frac{iq\hbar}{4m^2}(\sdotp)(\sdotE).
\end{equation}
The hermitian part of $A_2$ is given by
$A^H_2=(A_2+A^{\dag}_2)/2$,
\begin{equation}\label{H2FW_2}
A^H_2=-\frac{(\sdotp)^4}{8m^3}-\frac{iq\hbar}{8m^2}[\sdotp,\sdotE].
\end{equation}
Using $\sigma_i\sigma_j=\delta_{ij}+i\epsilon_{ijk}\sigma_k$, we
have
$[\sdotp,\sdotE]=-i\hbar\nabla\cdot\mathbf{E}+i\boldsigma\cdot\boldpi\times\mathbf{E}-i\boldsigma\cdot\mathbf{E}\times\boldpi$.
For static case, we have
$\boldpi\times\mathbf{E}=-\mathbf{E}\times\boldpi$. Therefore, up
to the second order of magnetic field, Eq.~(\ref{H2FW_2}) becomes
\begin{equation}\label{H2FW_3}
\begin{split}
H^{(2)}_{\mathFW}&=-\frac{\boldpi^4}{8m^3c^2}+\frac{q\hbar}{8m^3c^3}\left[\boldpi^2(\sdotB)+(\sdotB)\boldpi^2\right]\\
&~~-\frac{q\hbar}{4m^2c^2}\sep-\frac{q\hbar^2}{8m^2c^2}\nabla\cdot\mathbf{E}.
\end{split}
\end{equation}
The first term of Eq.~(\ref{H2FW_3}) is the relativistic
correction to the kinetic energy. The second term of
Eq.~(\ref{H2FW_3}) is the relativistic correction to the Zeeman
energy. The fourth and fifth terms of Eq.~(\ref{H2FW_3}) are the
spin-orbit interaction and the Darwin term which provides
heuristic evidence of the Zitterbewegung phenomenon
\cite{Darwin1928}. Combining Eqs.\ (\ref{H0FW_2}) and
(\ref{H2FW_3}), we obtain the Foldy-Wouthuysen transformed Dirac
Hamiltonian up to terms with $(\boldpi/mc)^4$:
\begin{equation}\label{HFW02}
\begin{split}
H_{\mathFW}&=mc^2+H^{(0)}_{\mathFW}+H^{(2)}_{\mathFW}\\
&=mc^2+V+\frac{\boldpi^2}{2m}-\frac{q\hbar}{2mc}\sdotB-\frac{\boldpi^4}{8m^3c^2}\\
&~~+\frac{q\hbar}{8m^3c^3}\left[\boldpi^2(\sdotB)+(\sdotB)\boldpi^2\right]\\
&~~-\frac{q\hbar^2}{8m^2c^2}\nabla\cdot\mathbf{E}-\frac{q\hbar}{4m^2c^2}\sep.
\end{split}
\end{equation}
Equation (\ref{HFW02}) is in agreement with the earlier
results \cite{Foldy1950, Froh1993} which are obtained by the standard FW
method.

If we take into account the terms of the second order in
electromagnetic fields, our result gives
$-\frac{e^2\hbar^2}{8m^3c^4}\mathbf{B}^2$. However, the FW
diagonalization method shows that the terms of the second order in
electromagnetic field should be
$\frac{e^2\hbar^2}{8m^3c^4}(\mathbf{E}^2-\mathbf{B}^2)$.
In comparison with the standard FW transformation method, this discrepancy suggests that the assumption that the series expansion of $X$ [Eq.~(\ref{SeriesX})] exists is
valid only in the low-energy and weak-field limit or in the
absence of an electric field.

In the following sections, we will obtain the FW transformed Dirac and Dirac-Pauli Hamiltonians in the low-energy and weak-field limit.

\section{FW transformed Dirac Hamiltonian}\label{sec:HFW}

The previous section shows that the Kutzelnigg diagonalization
method is valid when we consider only terms with linear
electromagnetic fields. We focus only on linear terms of
electromagnetic fields in comparison with the T-BMT equation. In
this section, we consider the static and homogeneous
electromagnetic field and neglect the product of fields in the FW
transformed Dirac Hamiltonian. The FW transformed Dirac
Hamiltonian contains the Dirac energy operator and Dirac string
operator [see Eqs.\ (\ref{EqExpandH}) and (\ref{EqS})]. We will
calculate $H_{\mathFW}^{(\ell)}$ from $\ell=0$ to $\ell=12$.
Nevertheless, we have to emphasize that Eq.~(\ref{EqA}) implies
that the $\ell$th order of Dirac energy operator is obtained from
the next order of the Dirac generating operator. Therefore, we
have to obtain the term of the generating operator up to the order
of $1/c^{13}$, i.e. $X_{13}$. The explicit forms of the expanding
terms of the generating operators can be derived by using Eqs.\
(\ref{App:EqX1}) and (\ref{App:EqXoe}). Up to the order of
$1/c^{13}$, we have
\begin{widetext}
\begin{equation}\label{App:SolveX}
\begin{split}
&X_1=\frac{\sdotp}{2m},~X_3=-\frac{1}{4}\frac{T}{m^2}\sdotp-\frac{1}{4}\frac{iq\hbar}{m^2}\boldsigma\cdot\mathbf{E},~X_5=\frac{1}{4}\frac{T^2}{m^3}\sdotp+\frac{3}{16}\frac{iq\hbar}{m^4}\boldpi^2(\sdotE)+\frac{1}{8}\frac{iq\hbar}{m^4}(\mathbf{E}\cdot\boldpi)(\sdotp),\\
&X_7=-\frac{5}{16}\frac{T^3}{m^4}\sdotp-\frac{5}{32}\frac{iq\hbar}{m^6}\boldpi^4(\sdotE)-\frac{3}{16}\frac{iq\hbar}{m^6}\boldpi^2(\mathbf{E}\cdot\boldpi)(\sdotp),\\
&X_9=\frac{7}{16}\frac{T^4}{m^5}\sdotp+\frac{35}{256}\frac{iq\hbar}{m^8}\boldpi^6(\sdotE)+\frac{29}{128}\frac{iq\hbar}{m^8}\boldpi^4(\mathbf{E}\cdot\boldpi)(\sdotp),\\
&X_{11}=-\frac{21}{32}\frac{T^5}{m^6}\sdotp-\frac{63}{1024}\frac{iq\hbar}{m^{10}}\boldpi^8(\sdotE)-\frac{65}{256}\frac{iq\hbar}{m^{10}}\boldpi^6(\mathbf{E}\cdot\boldpi)(\sdotp),\\
&X_{13}=\frac{33}{32}\frac{T^6}{m^7}\sdotp+\frac{231}{2048}\frac{iq\hbar}{m^{12}}\boldpi^{10}(\sdotE)+\frac{281}{1024}\frac{iq\hbar}{m^{12}}\boldpi^{8}(\mathbf{E}\cdot\boldpi)(\sdotp),
\end{split}
\end{equation}
\end{widetext}
where $T=(\sdotp)^2/2m$. The forms of $X_1$ and $X_3$ in
Eq.~(\ref{App:SolveX}) are also valid for inhomogeneous fields.
The expanding terms of the Dirac generating operator from $X_5$ to
$X_{13}$ in Eq.~(\ref{App:SolveX}) are valid only for homogeneous
fields. Inserting Eqs.~(\ref{App:SolveX}) into Eq.~(\ref{EqA}), we
can obtain each order of the Dirac energy operator. Furthermore,
we rewrite each order of the Dirac energy operator $A_{\ell}$ as
the combination of the hermitian part ($A^H_{\ell}$) and
anti-hermitian part ($A^{N}_{\ell}$),
\begin{equation}\label{EqAHN}
A_{\ell}=A^{H}_{\ell}+A^{N}_{\ell},
\end{equation}
where $A^H_{\ell}$ and $A^N_{\ell}$ satisfy
$A^{H\dag}_{\ell}=A^H_{\ell}$ and $A^{N\dag}_{\ell}=-A^N_{\ell}$.

The hermitian parts of the Dirac energy operator from $A^H_0$ to
$A^H_{12}$ are given by
\begin{equation}\label{EqAsH}
\begin{split}
&A^H_0=T+V,~A^H_2=-\frac{T^2}{2m}-\frac{q\hbar}{4m^2}\sep,\\
&A^H_4=\frac{T^3}{2m^2}+\frac{3}{16}\frac{q\hbar}{m^4}\boldpi^2\sep,\\
&A^H_6=-\frac{5}{8}\frac{T^4}{m^3}-\frac{5}{32}\frac{q\hbar}{m^6}\boldpi^4\sep,\\
&A^H_8=\frac{7}{8}\frac{T^5}{m^4}+\frac{35}{256}\frac{q\hbar}{m^8}\boldpi^6\sep,\\
&A^H_{10}=-\frac{21}{16}\frac{T^6}{m^5}-\frac{63}{512}\frac{q\hbar}{m^{10}}\boldpi^8\sep,\\
&A^H_{12}=\frac{33}{16}\frac{T^7}{m^6}+\frac{231}{2048}\frac{q\hbar}{m^{12}}\boldpi^{10}\sep,\\
\end{split}
\end{equation}
The anti-hermitian parts of the Dirac energy operator from $A^N_0$
to $A^N_{12}$ are given by
\begin{equation}\label{EqAsN}
\begin{split}
&A^N_0=0,~A^N_2=-\frac{iq\hbar}{4m^2}\mathbf{E}\cdot\boldpi,~A^N_4=+\frac{5}{16}\frac{iq\hbar}{m^4}\boldpi^2\mathbf{E}\cdot\boldpi,\\
&A^N_6=-\frac{11}{32}\frac{iq\hbar}{m^6}\boldpi^4\mathbf{E}\cdot\boldpi,~A^N_8=+\frac{93}{256}\frac{iq\hbar}{m^8}\boldpi^6\mathbf{E}\cdot\boldpi,\\
&A^N_{10}=-\frac{193}{512}\frac{iq\hbar}{m^{10}}\boldpi^8\mathbf{E}\cdot\boldpi,~A^N_{12}=+\frac{793}{2048}\frac{iq\hbar}{m^{12}}\boldpi^{10}\mathbf{E}\cdot\boldpi.\\
\end{split}
\end{equation}
We emphasize that  the second and higher order of electromagnetic
field will be neglected in Eqs.\ (\ref{EqAsH}) and (\ref{EqAsN}).
In order to simplify the present expression, the form of
$A_{\ell}$ still contains terms with non-linear electromagnetic
fields because the operator $T$ can be written as
$T=(\sdotp)^2/2m=\frac{1}{2m}(\boldpi^2-\frac{q\hbar}{c}\sdotB)$.
We will neglect these higher order terms when constructing
Hamiltonian. On the other hand, to evaluate the Dirac string
operator, we have to obtain the Dirac exponent operator by
expanding $\ln(1+X^{\dag}X)$. After straightforward calculations,
the expanding terms of the Dirac exponent operator $G_{\ell}$ (up
to $1/c^{12}$) are given by
\begin{equation}\label{EqGs}
\begin{split}
&G_2=\frac{T}{2m},~G_4=-\frac{5}{8}\frac{T^2}{m^2}-\frac{1}{4}\frac{q\hbar}{m^3}\sep,\\
&G_6=\frac{11}{12}\frac{T^3}{m^3}+\frac{5}{16}\frac{q\hbar}{m^5}\boldpi^2\sep,\\
&G_8=-\frac{93}{64}\frac{T^4}{m^4}-\frac{11}{32}\frac{q\hbar}{m^7}\boldpi^4\sep,\\
&G_{10}=\frac{193}{80}\frac{T^5}{m^5}+\frac{93}{256}\frac{q\hbar}{m^9}\boldpi^6\sep,\\
&G_{12}=-\frac{793}{192}\frac{T^6}{m^6}-\frac{193}{512}\frac{q\hbar}{m^{11}}\boldpi^8\sep.
\end{split}
\end{equation}
Since we always neglect terms with $E^2$, $B^2$, $EB$ and multiple
products of them, the kinetic energy operator $T$ commutes with
$\boldpi^{2k}\sep$ and $\boldpi^{2k}\mathbf{E}\cdot\boldpi$, and
we have
\begin{equation}\label{T comm}
\begin{split}
&[T,\boldpi^{2k}\sep]=0+o(f^2),\\
&[T,\boldpi^{2k}\mathbf{E}\cdot\boldpi]=0+o(f^2),\\
&[\boldpi^{2k}\sep,\boldpi^{2n}\sep]=0+o(f^2),
\end{split}
\end{equation}
where $o(f^2)$ represents the second order and higher orders of
homogeneous electromagnetic fields.

Applying Eqs.\ (\ref{EqAsH}), (\ref{EqAsN}), (\ref{EqGs}) and
(\ref{T comm}) to the Dirac string operators [Eq.~(\ref{EqS})], we
find that all the non-vanishing Dirac string operators $S_{\ell}$
(from $\ell=2$ to $\ell=12$) are proportional to second and higher
orders of electric and magnetic fields which are being neglected.
This can also be proved as follows.

Firstly, consider the Dirac string operator with only one Dirac
exponent operator, $S=[G,A^N]/2+o(G^2)$. The Dirac exponent
operator is $G=\sum_{\ell}G_{\ell}/c^{\ell}=G_T+G_{\mathrm{so}}$,
where $G_T$ is the term with collections of the kinetic energy
operator $T$, i.e.,
$G_T=T/2mc^2-(5/8)T^2/m^2c^4+(11/12)T^3/m^3c^6+\cdots$, and
$G_{\mathrm{so}}=(-1/4m^3c^4+5\boldpi^2/16m^5c^6-11\boldpi^4/32m^7c^8+\cdots)\hbar\sep=F(\boldpi^2)\sep$,
where $F(\boldpi^2)$ represents the power series of $\boldpi^2$.
The anti-hermitian part of the Dirac energy operator is
$A^N=\sum_{\ell}A^{N}_{\ell}/c^{\ell}=(-1/4m^2c^2+5\boldpi^2/16m^4c^4-11\boldpi^4/32m^6c^6+\cdots)i\hbar\mathbf{E}\cdot\boldpi=g(\boldpi^2)\mathbf{E}\cdot\boldpi$,
where $g(\boldpi^2)$ represents the power series of $\boldpi^2$.
Therefore, $[G,A^N]$ can be written as
$[G,A^N]=[G_T,A^N]+[G_{\mathrm{so}},A^N]$. Since we have
$[T,\boldpi^{2k}]=[T,\mathbf{E}\cdot\boldpi]=0+o(f^2)$, thus
$[G_T,A^N]=0+o(f^2)$. The commutator
$[G_{\mathrm{so}},A^N]=[F(\boldpi^2)\sep,g(\boldpi^2)\mathbf{E}\cdot\boldpi]$
also vanishes up to second-order terms of homogeneous
electromagnetic field, because we have
$[\sep,g(\boldpi^2)]=0+o(f^2)$,
$[\sep,\mathbf{E}\cdot\boldpi]=0+o(f^2)$,
$[F(\boldpi^2),\mathbf{E}\cdot\boldpi]=0+o(f^2)$ and
$[F(\boldpi^2),g(\boldpi^2)]=0$. We obtain $[G,A^N]=0+o(f^2)$, and
thus, the terms containing odd numbers of $G$ in the Dirac string
operator [see Eq.~(\ref{EqS NE})] always vanishes up to
second-order terms of homogeneous electromagnetic fields.

Secondly, consider the term containing two Dirac exponent
operators in the Dirac string operator, $[G,[G,A^H]]$. The
hermitian part of the Dirac energy operator can be written as
$A^H=\sum_{\ell}A^{\ell}/c^{\ell}=V+A^H_T+A^H_{\mathrm{so}}$,
where $A^H_{T}=T-T^2/2mc^2+T^3/2m^2c^4-5T^4/8m^3c^6+\cdots$ and
$A^H_{\mathrm{so}}=K(\boldpi^2)\sep$, where $K(\boldpi^2)$ is the
power series of $\boldpi^2$. The commutator $[G,A^H]$ becomes
$[G,A^H]=[G_T,V]+[G_T,A^H_T]+[G_T,A^{H}_{\mathrm{so}}]+[G_{\mathrm{so}},V]+[G_{\mathrm{so}},A^H_T]+[G_{\mathrm{so}},A^H_{\mathrm{so}}]$.
Since we have $[T,\boldpi^{2k}\sep]=0+o(f^2)$ and
$[\boldpi^{2k}\sep,\boldpi^{2n}\sep]=0+o(f^2)$, the commutators
$[G_T,A^{H}_{\mathrm{so}}]$, $[G_{\mathrm{so}},A^H_T]$ and
$[G_{\mathrm{so}},A^H_{\mathrm{so}}]$ vanish up to second-order
terms of homogeneous electromagnetic fields as well as
$[G_{\mathrm{so}},V]$. For the commutator $[G_T,V]$, using
$[T,V]=-i\hbar\mathbf{E}\cdot\boldpi/m$, we find that
$[G_T,V]=R(\boldpi^2)\mathbf{E}\cdot\boldpi+o(f^2)$, where
$R(\boldpi^2)$ is the power series of $\boldpi^2$. That is,
$[G,A^H]=R(\boldpi^2)\mathbf{E}\cdot\boldpi+o(f^2)$. Similar to
the commutator $[G,A^N]$, where
$A^N=g(\boldpi^2)\mathbf{E}\cdot\boldpi$, we find that this
implies that $[G,[G,A^H]]=0+o(f^2)$. Therefore, the terms
containing even numbers of the Dirac exponent operators in the
Dirac string operator [see Eq.~(\ref{EqS NE})] always vanishes up
to second-order terms of homogeneous electromagnetic fields. In
short, it can be shown that from $\ell=0$ to $\ell=12$, the
expanding terms of the Dirac string operator satisfies
\begin{equation}\label{EqVanS}
S_{\ell}=0+o(f^2),
\end{equation}
where $o(f^2)$ represents the second and higher orders of
electromagnetic fields.

Consider Eq.~(\ref{EqExpandH}) together with (\ref{EqVanS}), we
find that $H^{(\ell)}_{FW}$ is exactly equal to $A^H_{\ell}$, i.e.
\begin{equation}\label{EqHFW}
c^{\ell}H_{FW}^{(\ell)}=A^H_{\ell}+o(f^2),
\end{equation}
where $\ell=0,2,4,\cdots,12$. Equation (\ref{EqHFW}) is the main
result of this paper. This implies that the FW transformed Dirac
Hamiltonian is only determined by the hermitian part of the Dirac
energy operator regardless of the Dirac exponent operator $G$. We
have shown that Eq.~(\ref{EqHFW}) is valid at least up to
$1/c^{12}$. We believe that this result is valid to all higher
orders of $1/c$. Equation (\ref{EqHFW}) enables us to solely focus
on the hermitian part of the Dirac energy operator since the
anti-hermitian part can be exactly cancelled by the remaining
string operators. As a consequence, this result provides us a
method to obtain higher order terms faster than traditional
Foldy-Wouthuysen transformation.

Comparing the form of the resulting FW transformed Dirac
Hamiltonian with the classical Hamiltonian, we define the magnetic
moment $\boldmu$ and the scaled kinetic momentum $\boldxi$ as
\begin{equation}\label{def}
\boldmu=\frac{q\hbar}{2mc}\boldsigma,
\qquad
\boldxi=\frac{\boldpi}{mc}.
\end{equation}
On the other hand, the kinetic energy operator $T$ in
Eq.~(\ref{EqAsH}) can be replaced by
$T=\frac{\boldpi^2}{2m}-\frac{q\hbar}{2mc}\sdotB$, and after
neglecting second and higher orders of electromagnetic fields, the
FW transformed Hamiltonian [Eq.~(\ref{EqHFW})] becomes
\begin{widetext}
\begin{equation}\label{HFWs}
\begin{split}
H^{(0)}_{\mathFW}&=V+\frac{1}{2}mc^2\boldxi^2-\mudotB,\\
H^{(2)}_{\mathFW}&=-\frac{1}{8}mc^2\boldxi^4+\frac{1}{2}\boldxi^2\mudotB-\frac{1}{2}\mex,\\
H^{(4)}_{\mathFW}&=+\frac{1}{16}mc^2\boldxi^6-\frac{3}{8}\boldxi^6\mudotB+\frac{3}{8}\boldxi^4\mex,\\
H^{(6)}_{\mathFW}&=-\frac{5}{128}mc^2\boldxi^8+\frac{5}{16}\boldxi^4\mudotB-\frac{5}{16}\boldxi^4\mex,\\
H^{(8)}_{\mathFW}&=+\frac{7}{256}mc^2\boldxi^{10}-\frac{35}{128}\boldxi^8\mudotB+\frac{35}{128}\boldxi^6\mex,\\
H^{(10)}_{\mathFW}&=-\frac{21}{1024}mc^2\boldxi^{12}+\frac{63}{256}\boldxi^{10}\mudotB-\frac{63}{256}\boldxi^8\mex,\\
H^{(12)}_{\mathFW}&=+\frac{33}{2048}mc^2\boldxi^{14}+\frac{231}{1024}\boldxi^{12}\mudotB-\frac{231}{1024}\boldxi^{10}\mex.
\end{split}
\end{equation}
\end{widetext}
After substituting Eq.~(\ref{HFWs}) into Eq.~(\ref{HFW sum}), the
FW transformed Dirac Hamiltonian becomes a sum of two terms:
\begin{equation}\label{HFW OandS}
\begin{split}
H_{\mathFW}&=\sum_{\ell}H^{(\ell)}_{\mathFW}\\
&=\Horbit+\Hspin,
\end{split}
\end{equation}
where the orbital Hamiltonian $\Horbit$ is the kinetic energy
(including the rest mass energy) plus the potential energy,
\begin{equation}\label{EqHo}
\begin{split}
\Horbit&=mc^2(1+\frac{1}{2}\boldxi^2-\frac{1}{8}\boldxi^4+\frac{1}{16}\boldxi^6-\frac{5}{128}\boldxi^8\\
&~~+\frac{7}{256}\boldxi^{10}-\frac{21}{1024}\boldxi^{12}+\frac{33}{2048}\boldxi^{14})+V,
\end{split}
\end{equation}
and the spin Hamiltonian $\Hspin$ is the Hamiltonian of intrinsic
magnetic moment in electromagnetic fields,
\begin{equation}\label{EqHs}
\begin{split}
\Hspin&=-(1-\frac{1}{2}\boldxi^2+\frac{3}{8}\boldxi^4-\frac{5}{16}\boldxi^6+\frac{35}{128}\boldxi^8-\frac{63}{256}\boldxi^{10}\\
&~~+\frac{231}{1024}\boldxi^{12})\mudotB+(-\frac{1}{2}+\frac{3}{8}\boldxi^2-\frac{5}{16}\boldxi^4+\frac{35}{128}\boldxi^6\\
&~~-\frac{63}{256}\boldxi^8+\frac{231}{1024}\boldxi^{10})\mex.\\
\end{split}
\end{equation}
In the following section, we will show that the FW transformed
Dirac Hamiltonian is equivalent to the Hamiltonian obtained from
T-BMT equation with $g=2$.

\section{FW transformed Dirac Hamiltonian and classical Hamiltonian}\label{sec:TBMT}

The orbital Hamiltonian $\Horbit$ [Eq.~(\ref{EqHo})] is expected
to be equivalent to the classical relativistic energy $\gamma
mc^2+V$. However, the boost velocity in T-BMT equation is not
$\boldxi$ \cite{TWChen2010}. Take the series expansion of
$(1+\boldxi^2)^{1/2}$ into account,
\begin{equation}\label{series_xi}
\begin{split}
(1+\boldxi^2)^{1/2}&=1+\frac{1}{2}\boldxi^2-\frac{1}{8}\boldxi^4+\frac{1}{16}\boldxi^6-\frac{5}{128}\boldxi^8\\
&~~+\frac{7}{256}\boldxi^{10}-\frac{21}{1024}\boldxi^{12}+\frac{33}{2048}\boldxi^{14}\\
&~~-\frac{429}{32768}\boldxi^{16}+\cdots,
\end{split}
\end{equation}
we find that the series of $\boldxi^2$ in Eq.~(\ref{EqHo}) is
exactly equal to Eq.~(\ref{series_xi}) up to $\boldxi^{14}$. This
enable us to define the boost operator $\boost$ via the Lorentz
operator $\gammaop$,
\begin{equation}\label{xiboost}
(1+\boldxi^2)^{1/2}=\gammaop=\frac{1}{\sqrt{1-\boost^2}}.
\end{equation}
In this sense, the orbital Hamiltonian can now be written as
$\Horbit=\gammaop mc^2+V$. In classical relativistic theory, the
Lorentz factor $\gamma$ is related to boost velocity by
$\gamma=1/\sqrt{1-\beta^2}$. However, in the relativistic quantum
mechanics since different components of the kinetic momentum
operator $\boldpi$ do not commute with one another, the boost
operator $\boost$ should not simply satisfy the form
$\gammaop=1/\sqrt{1-\boost^2}$. We will go back to this point when
discussing the spin Hamiltonian. The boost operator $\boost$ plays
an important role on showing the agreement between the spin
Hamiltonian $\Hspin$ and the T-BMT equation.

The spin Hamiltonian can be written as a sum of Zeeman Hamiltonian
$\Hze$ and spin-orbit interaction $\Hso$,
\begin{equation}
\Hspin=\Hze+\Hso.
\end{equation}
The Zeeman Hamiltonian $\Hze$ is the relativistic correction to
the Zeeman energy:
\begin{equation}\label{Hze}
\begin{split}
\Hze&=-(1-\frac{1}{2}\boldxi^2+\frac{3}{8}\boldxi^4-\frac{5}{16}\boldxi^6+\frac{35}{128}\boldxi^8-\frac{63}{256}\boldxi^{10}\\
&~~+\frac{231}{1024}\boldxi^{12})\mudotB.
\end{split}
\end{equation}
The spin-orbit interaction $\Hso$ is the interaction of electric
field and the electric dipole moment arising from the boost on the
intrinsic spin magnetic moment:
\begin{equation}\label{Hso}
\begin{split}
\Hso&=-(\frac{1}{2}-\frac{3}{8}\boldxi^2+\frac{5}{16}\boldxi^4-\frac{35}{128}\boldxi^6\\
&~~+\frac{63}{256}\boldxi^8-\frac{231}{1024}\boldxi^{10})\mex.
\end{split}
\end{equation}
We first focus on the series in the Zeeman Hamiltonian. Consider
the series expansion of $(1+\boldxi^2)^{-1/2}$,
\begin{equation}\label{Series1}
\begin{split}
(1+\boldxi^2)^{-1/2}&=1-\frac{1}{2}\boldxi^2+\frac{3}{8}\boldxi^4-\frac{5}{16}\boldxi^6+\frac{35}{128}\boldxi^8\\
&~~-\frac{63}{256}\boldxi^{10}+\frac{231}{1024}\boldxi^{12}-\frac{429}{2048}\boldxi^{14}+\cdots,
\end{split}
\end{equation}
we find that the series in $\Hze$ is exactly equal to
$(1+\boldxi^2)^{-1/2}$ up to $\boldxi^{12}$. Therefore, the Zeeman
Hamiltonian Eq.~(\ref{Hze}) can be written as
\begin{equation}\label{Hze_boost}
\Hze=-\frac{1}{\gammaop}\mudotB.
\end{equation}
On the other hand, the the spin-orbit term in the T-BMT
Hamiltonian transforms like $[g/2-\gamma/(1+\gamma)]$ and $g=2$
for the Dirac Hamiltonian. Therefore, consider the series
expansion of $(1-\gammaop/(1+\gammaop))(1/\gammaop)$, we have
\begin{equation}\label{Series2}
\begin{split}
\left(1-\frac{\gammaop}{1+\gammaop}\right)\frac{1}{\gammaop}&=\frac{1}{\sqrt{1+\boldxi^2}}-\frac{1}{1+\sqrt{1+\boldxi^2}}\\
&=\frac{1}{2}-\frac{3}{8}\boldxi^2+\frac{5}{16}\boldxi^4-\frac{35}{128}\boldxi^6\\
&~~+\frac{63}{256}\boldxi^8-\frac{231}{1024}\boldxi^{10}+\frac{429}{2048}\boldxi^{12}+\cdots,
\end{split}
\end{equation}
where $\gammaop(1+\gammaop)^{-1}=(1+\gammaop)^{-1}\gammaop$ was
used.\footnote{The identity can be shown as follows.
$\gammaop(1+\gammaop)^{-1}=[(1+\gammaop)\gammaop^{-1}]^{-1}=(\gammaop^{-1}+1)^{-1}=[\gammaop^{-1}(1+\gammaop)]^{-1}=(1+\gammaop)^{-1}\gammaop$.}
The series in Eq.~(\ref{Hso}) is in agreement with Eq.~(\ref{Series2}) up to $\boldxi^{10}$. Therefore, we have
\begin{equation}\label{Hso_boost}
\Hso=-\left(1-\frac{\gammaop}{1+\gammaop}\right)\frac{1}{\gammaop}\mex.
\end{equation}
We note that if Eq.~(\ref{Hso_boost}) is in complete agreement
with the T-BMT equation, the boost velocity operator $\boost$ must
be defined by $\boost=\frac{1}{\gammaop}\boldxi$. In general, the
commutator $[\boldxi,1/\gammaop]$ is not equal to zero, and
Eq.~(\ref{xiboost}) cannot be satisfied. However, since we require
that the FW transformed Dirac Hamiltonian $H_{\mathFW}$ is linear
in electromagnetic fields, the magnetic field obtained from the
operator $[\boldxi,1/\gammaop]$ should be neglected. In that
sense, the commutator $[\boldxi,1/\gammaop]$ should be identified
as zero in this case, and the boost operator can be written as
\begin{equation}\label{xiboost1}
\boost=\frac{1}{\gammaop}\boldxi=\boldxi\frac{1}{\gammaop}.
\end{equation}
It can be shown that Eq.~(\ref{xiboost1}) satisfies Eq.~(\ref{xiboost}). Therefore, the spin Hamiltonian Eq.~(\ref{EqHs})
with substitutions of Eqs.~(\ref{Hze_boost}) and (\ref{Hso_boost})
becomes
\begin{equation}\label{Hspin}
\begin{split}
\Hspin&=\Hze+\Hso\\
&=-\boldmu\cdot\left[\frac{1}{\gammaop}\mathbf{B}-\left(1-\frac{\gammaop}{1+\gammaop}\right)\boost\times\mathbf{E}\right].
\end{split}
\end{equation}
Up to the twentieth order there is a complete agreement between
the spin part of the FW transformed Dirac Hamiltonian and the
T-BMT equation with $g=2$. The FW transformed Hamiltonian is given
by
\begin{equation}\label{HFW total}
\begin{split}
H_{\mathFW}&=\Horbit+\Hspin\\
&=V+\gammaop
mc^2-\boldmu\cdot\left[\frac{1}{\gammaop}\mathbf{B}-\left(1-\frac{\gammaop}{1+\gammaop}\right)\boost\times\mathbf{E}\right],
\end{split}
\end{equation}
which is in agreement with the classical Hamiltonian with $g=2$.
In the next section, we take into account the Pauli anomalous
magnetic moment and show that the classical correspondence of the
Dirac-Pauli Hamiltonian is the classical Hamiltonian with
$g\neq2$.

\section{FW transformation for Dirac-Pauli Hamiltonian}\label{sec:TBMT2}

In the previous section, the agreement to the classical
Hamiltonian is shown to be complete up to terms of the order
$(\boldpi/mc)^{14}$ in the absence of anomalous electron magnetic
moment, i.e., $g=2$. The Dirac electron including the Pauli
anomalous magnetic moment can be described by the Dirac-Pauli
Hamiltonian denoted by $\mathcal{H}$ which contains the Dirac
Hamiltonian as well as anomalous magnetic interaction $V_B$ and
anomalous electric interaction $V_E$,
\begin{equation}
\begin{split}
\mathcal{H}&=H_D+\left(\begin{array}{cc} V_B&iV_E\\
-iV_E&-V_B
\end{array}\right)\\
&=\left(\begin{array}{cc}
H_+&H_0\\
H_0^{\dag}&H_-
\end{array}\right)
\end{split}
\end{equation}
where $H_+=V+V_B+mc^2$, $H_-=V-V_B-mc^2$ and $H_0=h_0+iV_E$. The
Dirac Hamiltonian $H_D$ is given in Eq.~(\ref{Dirac}), and
\begin{equation}
V_B=-\mu'\sdotB,~V_E=\mu'\sdotE.
\end{equation}
The coefficient $\mu'$ is defined as
\begin{equation}
\mu'=\left(\frac{g}{2}-1\right)\frac{q\hbar}{2mc}.
\end{equation}
For an electron with $g=2$, we have $V_B=0$ and $V_E=0$. Applying
the unitary transformation Eq.~(\ref{U})
\begin{equation}\label{UHDPU}
U=\left(\begin{array}{cc} \mathcal{Y}&\mathcal{Y}\mathcal{X}^{\dag}\\
-\mathcal{Z}\mathcal{X}&\mathcal{Z}
\end{array}\right)
\end{equation}
to the Dirac-Pauli Hamiltonian, the self-consistent equation for
the Dirac-Pauli generating operator $\mathcal{X}$ is given by the
requirement of vanishing off-diagonal term of
$U\mathcal{H}U^{\dag}$, i.e.
\begin{equation}\label{DP X}
\begin{split}
2mc^2\mathcal{X}&=[V,\mathcal{X}]+h_0-\mathcal{X}h_0\mathcal{X}\\
&~~-iV_E-i\mathcal{X}V_E\mathcal{X}-\{\mathcal{X},V_B\},
\end{split}
\end{equation}
where $h_0=c\sdotp$. The FW transformed Dirac-Pauli Hamiltonian
can be obtained from the upper-left block diagonal term of
$U\mathcal{H}U^{\dag}$, and it is given by
\begin{equation}\label{HFWDP0}
\mathcal{H}_{\mathFW}=
\mathcal{Y}\left(H_++\mathcal{X}^{\dag}H_0^{\dag}
+H_0\mathcal{X}+\mathcal{X}^{\dag}H_-\mathcal{X}\right)\mathcal{Y}.
\end{equation}
Similar to the derivation of Eq.~(\ref{HFW}), we find that the FW
transformed Dirac-Pauli Hamiltonian [Eq.~(\ref{HFWDP0})] can also
be simplified as (see Appendix~\ref{App:Ham2})
\begin{equation}\label{H FWDP}
\mathcal{H}_{\mathFW}=mc^2+e^{\mathcal{G}/2}\mathcal{A}e^{-\mathcal{G}/2},
\end{equation}
where the Dirac-Pauli energy operator $\mathcal{A}$ and the
Pauli-Dirac exponent operator $\mathcal{G}$ are given by
\begin{equation}\label{DP AG}
\begin{split}
&\mathcal{A}=V+h_0\mathcal{X}+V_B+iV_E\mathcal{X},\\
&\mathcal{G}=\ln\left(1+\mathcal{X}^{\dag}\mathcal{X}\right).
\end{split}
\end{equation}
Similar to Eq.~(\ref{HFW NE}) obtained from the requirement of
hermiticity of $H_{\mathFW}$, we find that the FW transformed
Dirac-Pauli Hamiltonian also satisfies
$\mathcal{H}_{\mathFW}=\mathcal{H}_{\mathFW}^{\dag}=mc^2+e^{-\mathcal{G}/2}\mathcal{A}e^{\mathcal{G}/2}$
and the FW transformed Dirac-Pauli Hamiltonian can be rewritten as
\begin{equation}
\mathcal{H}_{\mathFW}=mc^2+\mathcal{A}^H+\mathcal{S},
\end{equation}
where $\mathcal{A}^H$ is the hermitian part of the Dirac-Pauli
energy operator and the Dirac-Pauli string operator $\mathcal{S}$
is the same as Eq.~(\ref{EqS NE}) by the replacements
$A\rightarrow\mathcal{A}$ and $G\rightarrow\mathcal{G}$, i.e.
\begin{equation}\label{EqS DP NE}
\begin{split}
\mathcal{S}&=\frac{1}{2}[\mathcal{G},\mathcal{A}^N]+\frac{1}{2!2^2}[\mathcal{G},[\mathcal{G},\mathcal{A}^H]]+\frac{1}{3!2^3}[\mathcal{G},[\mathcal{G},[\mathcal{G},\mathcal{A}^N]]]\\
&~~+\frac{1}{4!2^4}[\mathcal{G},[\mathcal{G},[\mathcal{G},[\mathcal{G},\mathcal{A}^H]]]]+\cdots.
\end{split}
\end{equation}
Similar to the Dirac string operator, the anti-hermitian part of
the Dirac-Pauli energy operator always appears in those terms with
odd numbers of Dirac-Pauli exponent operators, and the hermitian
part of the Dirac-Pauli energy operator always appears in those
terms with even numbers of Dirac exponent operators. The power
series solutions to the Dirac-Pauli generating operator can be
obtained by means of Eq.~(\ref{DP X}) via substitution of the
series expansion $\mathcal{X}=\sum_i\mathcal{X}_k/c^k$,
$k=1,2,3,\cdots$, and each order of Dirac-Pauli energy operator
$\mathcal{A}_k$ can be obtained from
$\mathcal{A}=\sum_k\mathcal{A}_k/c^k$ by using Eq.~(\ref{DP AG}).
Each order of Dirac-Pauli energy operators can be decomposed into
hermitian ($\mathcal{A}^H_k$) and anti-hermitian
($\mathcal{A}^N_k$) parts,
$\mathcal{A}_k=\mathcal{A}^H_k+\mathcal{A}^N_k$. As a result, the
FW transformed Dirac-Pauli Hamiltonian can be written as
\begin{equation}
\mathcal{H}_{FW}=mc^2+\sum_{k=0,1,2,\cdots}\mathcal{H}^{(k)}_{FW}
\end{equation}
with
\begin{equation}
c^k\mathcal{H}^{(k)}_{FW}=\mathcal{A}^H_{k}+\mathcal{S}_k.
\end{equation}
To obtain the FW transformed Dirac-Pauli Hamiltonian up to $k=12$,
the largest order of the Dirac-Pauli generating operator must have
the order of $k=13$, i.e., $\mathcal{X}_{13}$. This is because the
operator $h_0=c\boldsigma\cdot\boldpi$ is of the order of $c$, the
order of the Dirac-Pauli energy operator is lower than that of the
Dirac-Pauli generating operator. Furthermore, since each order of
the Dirac-Pauli generating operator must equal that of the Dirac
generating operator when $g=2$, we can rewrite the Dirac-Pauli
generating operator ($\mathcal{X}_k$) as the sum of the Dirac
generating operator ($X_k$) and the anomalous generating operator
($X_k'$), namely,
\begin{equation}\label{XX'}
\mathcal{X}_k=X_k+X_k'+o(f^2),
\end{equation}
where the anomalous generating operator $X_k'$ vanishes when
$g=2$. Similar to the derivation of power series solution to the
Dirac generating operator shown in the previous section, the
explicit forms of different orders of the anomalous generating
operator $X_k'$ are given by ($k=1,2,\cdots,13$)
\begin{widetext}
\begin{equation}\label{X'k}
\begin{split}
&X_1'=0,~X_2'=0,~X_3'=-\frac{i\mu''}{2m}\sdotE,~X_4'=\frac{\mu''}{2m^2}\mathbf{B}\cdot\boldpi,~X_5'=\frac{3}{8}\frac{i\mu''}{m^3}\boldpi^2(\sdotE)-\frac{i\mu''}{4m^3}(\sdotp)(\mathbf{E}\cdot\boldpi)\\
&X_6'=-\frac{3}{8}\frac{\mu''}{m^4}\boldpi^2(\mathbf{B}\cdot\boldpi),~X_7'=-\frac{5}{16}\frac{i\mu''}{m^5}\boldpi^4(\sdotE)+\frac{1}{4}\frac{i\mu''}{m^5}\boldpi^2(\sdotp)(\mathbf{E}\cdot\boldpi),\\
&X_8'=\frac{5}{16}\frac{\mu''}{m^6}\boldpi^4(\mathbf{B}\cdot\boldpi),~X_9'=\frac{35}{128}\frac{i\mu''}{m^7}\boldpi^6(\sdotE)-\frac{15}{64}\frac{i\mu''}{m^7}\boldpi^4(\sdotp)(\mathbf{E}\cdot\boldpi),\\
&X_{10}'=-\frac{35}{128}\frac{\mu''}{m^8}\boldpi^6(\mathbf{B}\cdot\boldpi),~X_{11}'=-\frac{63}{256}\frac{i\mu''}{m^9}\boldpi^8(\sdotE)+\frac{7}{32}\frac{i\mu''}{m^9}\boldpi^6(\sdotp)(\mathbf{E}\cdot\boldpi),\\
&X_{12}'=\frac{63}{256}\frac{\mu''}{m^{10}}\boldpi^8(\mathbf{B}\cdot\boldpi),~X_{13}'=\frac{231}{1024}\frac{i\mu''}{m^{11}}\boldpi^{10}(\sdotE)-\frac{105}{512}\frac{i\mu''}{m^{11}}\boldpi^8(\sdotp)(\mathbf{E}\cdot\boldpi),\\
\end{split}
\end{equation}
\end{widetext}
where $\mu''=(g/2-1)q\hbar/2m=c\mu'$. We note that since the
gyromagnetic ratio always accompanies linear-order terms of
electric or magnetic fields, the operator $X_k'$ is proportional
to electromagnetic fields and contains the kinetic momentum
operator.

Substituting Eq.~(\ref{XX'}) into Eq.~(\ref{DP AG}), the
Dirac-Pauli energy operator ($\mathcal{A}_k$) can be written as
the sum of the energy operator for the Dirac Hamiltonian ($A_k$)
and anomalous energy operator ($A'_k$),
\begin{equation}
\mathcal{A}_k=A_k+A'_k+o(f^2),
\end{equation}
where the expanding terms of the Dirac energy operator $A_k$ from
$k=0$ to $k=12$ are given in Eqs.~(\ref{EqAsH}) and (\ref{EqAsN}).
The $k$th order of the anomalous energy operator is related to
$k$th orders of Dirac generating operator and anomalous generating
operator by
\begin{equation}
\begin{split}
&A'_0=0,~A'_1=cV_B,\\
&A'_{k}=(h_0/c)X'_{k+1}+icV_EX_{k-1}, k=2,4,\cdots12,\\
&A'_{k}=(h_0/c)X_{k+1}, k=3,5,\cdots,11.
\end{split}
\end{equation}
Using Eqs.~(\ref{App:SolveX}) and (\ref{X'k}), the hermitian parts
of the expanding terms of the anomalous energy operator from
zeroth order to twentieth orders are given by
\begin{equation}\label{A'k H}
\begin{split}
&A'^H_0=0,~A'^H_1=-\mu''\sdotB,\\
&A'^H_2=-\frac{\mu''}{m}\sep,\\
&A'^H_3=\frac{1}{2}\frac{\mu''}{m^2}(\sdotp)(\Bdotp),\\
&A'^H_4=\frac{\mu''}{m^3}\boldpi^2\sep,\\
&A'^H_5=-\frac{3}{8}\frac{\mu''}{m^4}\boldpi^2(\sdotp)(\Bdotp),\\
&A'^H_6=-\frac{3}{8}\frac{\mu''}{m^5}\boldpi^4\sep,\\
&A'^H_7=\frac{5}{16}\frac{\mu''}{m^6}\boldpi^4(\sdotp)(\Bdotp),\\
&A'^H_8=\frac{5}{16}\frac{\mu''}{m^7}\boldpi^6\sep\\
&A'^H_9=-\frac{35}{128}\frac{\mu''}{m^8}\boldpi^6(\sdotp)(\Bdotp),\\
&A'^H_{10}=-\frac{35}{128}\frac{\mu''}{m^9}\boldpi^8\sep,\\
&A'^H_{11}=\frac{63}{256}\frac{\mu''}{m^{10}}\boldpi^8(\sdotp)(\Bdotp),\\
&A'^H_{12}=\frac{63}{256}\frac{\mu''}{m^{11}}\boldpi^{10}\sep.\\
\end{split}
\end{equation}
For example, consider the twentieth order of the anomalous
operator $A'_{12}$, which is given by
$A'_{12}=(h_0/c)X'_{13}+icV_EX_{11}$. Substituting $V_E$,
$X'_{13}$ and $X_{11}$ into $A'_{12}$ and neglecting the second
order of homogeneous electromagnetic fields, we find that
\begin{equation}
\begin{split}
A'_{12}&=\frac{231}{1024}\frac{i\mu''}{m^{11}}\boldpi^{10}(\sdotp)(\sdotE)-\frac{21}{32}\frac{i\mu''}{m^6}T^5(\sdotE)(\sdotp)\\
&~~-\frac{105}{512}\frac{i\mu''}{m^{11}}\boldpi^{10}\sdotE\\
&=\left(\frac{231}{1024}+\frac{21}{32}\times\frac{1}{32}\right)\frac{\mu''}{m^{11}}\boldpi^{10}\sep\\
&~~+\left(-\frac{105}{512}+\frac{231}{1024}-\frac{21}{32}\times\frac{1}{32}\right)\frac{i\mu''}{m^{11}}\boldpi^{10}\sdotE\\
&=\frac{63}{256}\frac{\mu''}{m^{11}}\boldpi^{10}\sep,
\end{split}
\end{equation}
where in the second equality we have used
$(\sdotE)(\sdotp)=\mathbf{E}\cdot\boldpi+i\sep$ and
$\mathbf{E}\times\boldpi=-\boldpi\times\mathbf{E}$ for homogeneous
fields, and the kinetic energy operator is replaced by
$T\rightarrow\boldpi^2/2m$. The anti-hermitian part of $A'_{12}$
is $i\mu''\boldpi^{10}\sdotE/m^{11}$ and its numerical coefficient
is zero. Interestingly, we find that all the anti-hermitian part
of $A'_k$ from $k=0$ to $k=12$ vanish up to second-order terms of
homogeneous electromagnetic fields, i.e.
\begin{equation}\label{A'k N}
A'^N_k=0+o(f^2).
\end{equation}
On the other hand, the series expansion of the Dirac-Pauli
exponent operator
$\mathcal{G}=\ln(1+\mathcal{X}^{\dag}\mathcal{X})$ can also be
written as $\mathcal{G}=\sum_{k}\mathcal{G}_k/c^k$ and
$\mathcal{G}_{k}=G_k+G'_k$, where  $G_k$ (the $k$th order of the
Dirac exponent operator) is given in Eq.~(\ref{EqGs}) and $G'_k$
is the $k$th order of the anomalous exponent operator. The
expanding terms of the anomalous exponent operators $G'_k$ from
$k=1$ to $k=12$ are as follows:
\begin{equation}\label{G'k}
\begin{split}
&G'_1=0,~G'_2=0,~G'_3=0,G'_4=-\frac{\mu''}{2m^2}\sep,\\
&G'_5=\frac{\mu''}{2m^3}(\sdotp)(\Bdotp), G'_6=\frac{5}{8}\frac{\mu''}{m^4}\boldpi^2\sep,\\
&G'_7=-\frac{5}{8}\frac{\mu''}{m^5}\boldpi^2(\sdotp)(\Bdotp),\\
&G'_8=-\frac{11}{16}\frac{\mu''}{m^6}\boldpi^4\sep,\\
&G'_9=\frac{11}{16}\frac{\mu''}{m^7}\boldpi^4(\sdotp)(\Bdotp),\\
&G'_{10}=\frac{93}{128}\frac{\mu''}{m^8}\boldpi^6\sep,\\
&G'_{11}=-\frac{93}{128}\frac{\mu''}{m^9}\boldpi^6(\sdotp)(\Bdotp),\\
&G'_{12}=-\frac{193}{256}\frac{\hbar}{m^{10}}\boldpi^8\sep,\\
\end{split}
\end{equation}
which all vanish when $g=2$ and each order of the anomalous
exponent operator must be proportional to electromagnetic fields.
Substitute Eqs.~(\ref{A'k H}), (\ref{A'k N}) and (\ref{G'k}) into
Eq.~(\ref{EqS DP NE}), it can be shown that similar to the result
of the Dirac string operator, the Dirac-Pauli string operator also
vanishes up to second-order terms of homogeneous electromagnetic
fields; i.e., we have
\begin{equation}
\mathcal{S}_{k}=0+o(f^2).
\end{equation}
This can be proved as follows.

Firstly, consider the term containing only one Dirac-Pauli
exponent operator in the Dirac-Pauli string operator [see
Eq.~(\ref{EqS DP NE})]. It is given by
$[\mathcal{G},\mathcal{A}^{N}]/2$. Since $\mathcal{G}=G+G'$ and
$\mathcal{A}^{N}=A^N+A'^N$, we have
$[\mathcal{G},\mathcal{A}^{H}]/2=[G,A^N]/2+[G,A'^N]/2+[G',A^N]/2+[G',A'^N]/2$,
where $[G,A^N]/2$ is the Dirac string operator containing only one
Dirac exponent operator and it has been shown that
$[G,A^N]/2=0+o(f^2)$. The anomalous exponent operator can be
written as
$G'=\sum_{k}G'_k/c^k=F_1(\boldpi^2)\sep+F_2(\boldpi^2)(\sdotp)(\Bdotp)$,
where $F_1(\boldpi^2)$ represents a power series of $\boldpi^2$ as
well as $F_2(\boldpi^2)$. We note that
$A'^N=\sum_{k}A'^N/c^k=0+o(f^2)$ [see Eq.~(\ref{A'k N})]. It is
obvious that the second term $[G,A'^N]$ and fourth term
$[G',A'^N]$ vanish up to second-order terms of homogeneous
electromagnetic fields. The third term
$[G',A^N]=[f_1(\boldpi^2)\sep+f_2(\boldpi^2)(\sdotp)(\Bdotp),g(\boldpi^2)\mathbf{E}\cdot\boldpi]$
also vanishes because we have
$[f_1(\boldpi^2),g(\boldpi^2)]=[f_2(\boldpi^2),g(\boldpi^2)]=0$
and
$[\sep,\mathbf{E}\cdot\boldpi]=[(\sdotp)(\Bdotp),\mathbf{E}\cdot\boldpi]=[\sep,g(\boldpi^2)]=[(\sdotp)(\Bdotp),g(\boldpi^2)]=[f_1(\boldpi^2),\mathbf{E}\cdot\boldpi]=[f_2(\boldpi^2),\mathbf{E}\cdot\boldpi]=0+o(f^2)$.

Therefore, we have $[\mathcal{G},\mathcal{A}^N]/2=0+o(f^2)$. Since
the commutator $[\mathcal{G},\mathcal{A}^N]/2$ always appears in
those terms with odd numbers of the Dirac-Pauli exponent operators
[see Eq.~(\ref{EqS DP NE})], this implies that the terms with odd
numbers of $\mathcal{G}$ always vanish up to second-order terms of
homogeneous electromagnetic fields.

Secondly, consider the terms with two Dirac-Pauli exponent
operators in the Dirac-Pauli string operator. It is given by
$[\mathcal{G},[\mathcal{G},\mathcal{A}^{H}]]/2!2^2=[G,[G,A^H]]/2!2^2+[G,[G,A'^H]]/2!2^2+[G,[G',A^H]]/2!2^2+[G',[G,A^H]]/2!2^2+o(f^2)$,
where we have neglected the second-order terms of electromagnetic
fields, such as $[G,[G',A'^H]]$, $[G',[G,A'^H]]$, $[G',[G',A^H]]$
and $[G',[G',A'^H]]$. The first term $[G,[G,A^H]]$ is the Dirac
string operator containing only two Dirac exponent operators and
it has been show that $[G,[G,A^H]]=0+o(f^2)$. The anomalous energy
operator can be written as
$A'^H=\sum_kA'^H_k/c^k=K_1(\boldpi^2)\sep+K_2(\boldpi^2)(\sdotp)(\Bdotp)$,
where $K_1(\boldpi^2)$ represents the power series of $\boldpi^2$
as well as $K_2(\boldpi^2)$. The commutator $[G,A'^H]$ can be
written as
$[G,A'^H]=[G_T+G_{\mathrm{so}},A'^H]=[G_T,A'^H]+[G_{\mathrm{so}},A'^H]$,
where $G_{\mathrm{so}}=F(\boldpi^2)\sep$ and
$G_T=T/2m-(5/8)T^2/m^2+(11/12)T^3/m^3+\cdots$. Using Eq.~(\ref{T
comm}), it can be shown that $[G_T,A'^H]=0+o(f^2)$ and
$[G_{\mathrm{so}},A'^H]=0+o(f^2)$, and thus, the second term
$[G,[G,A'^H]]$ vanishes up to second-order terms of homogeneous
electromagnetic fields. Consider the third term $[G,[G',A^H]]$,
where the commutator $[G',A^H]$ becomes
$[G',A^H]=[F_1(\boldpi^2)\sep+F_2(\boldpi^2)(\sdotp)(\Bdotp),V+A^H_T+A^H_{\mathrm{so}}]=[F_1(\boldpi^2)\sep,V]+[F_2(\boldpi^2)(\sdotp)(\Bdotp),V]+o(f^2)$.
However, the two commutators $[F_1(\boldpi^2)\sep,V]$ and
$[F_2(\boldpi^2)(\sdotp)(\Bdotp),V]$ are proportional to the
second order of homogeneous electromagnetic fields since
$[\boldpi,V]=iq\hbar\mathbf{E}$, and thus we have
$[G,[G',A^H]]=0+o(f^2)$. The fourth term $[G',[G,A^H]]$ also
vanishes up to $o(f^2)$ since it has been shown that
$[G,A^H]=R(\boldpi^2)\mathbf{E}\cdot\boldpi+o(f^2)$ and $G'$
contains first-order terms of homogeneous electromagnetic fields.

Therefore, we have shown that
$[\mathcal{G},[\mathcal{G},\mathcal{A}^{H}]]/2!2^2=0+o(f^2)$.
Since the commutator $[\mathcal{G},[\mathcal{G},\mathcal{A}^H]]$
always appears in those terms with even number of the Dirac-Pauli
exponent operator [see Eq.~(\ref{EqS DP NE})], this implies that
the terms with even numbers of $\mathcal{G}$ always vanishes up to
second order of homogeneous electromagnetic fields.

As a consequence, the FW transformed Dirac-Pauli Hamiltonian is
determined only by the hermitian part of Dirac-Pauli energy
operator, i.e.
\begin{equation}
c^k\mathcal{H}^{(k)}_{\mathFW}=\mathcal{A}_k^H+o(f^2).\\
\end{equation}
Since the Dirac-Pauli energy operator is composed of the Dirac
energy operator and anomalous energy operator,
$A^{H}_k=A^H_k+A'^H_k$, and the Dirac energy operator is related
to the FW transformed Dirac Hamiltonian by
$c^kH^{(k)}_{\mathFW}=A^H_k$, the FW transformed Dirac-Pauli
Hamiltonian can be written as the sum of the FW transformed Dirac
Hamiltonian and the anomalous Hamiltonian:
\begin{equation}
\mathcal{H}_{\mathFW}=H_{\mathFW}+H'_{\mathFW}.
\end{equation}
The $k$th order of the FW transformed Pauli-Dirac Hamiltonian can
be written as
\begin{equation}
\mathcal{H}^{(k)}_{\mathFW}=H^{(k)}_{\mathFW}+H'^{(k)}_{\mathFW},
\end{equation}
where the the $k$th order of the anomalous Hamiltonian
$H'_{\mathFW}$ denoted as $H'^{(k)}_{\mathFW}$ is determined by
the $k$th order of the anomalous energy operator:
\begin{equation}
c^kH'^{(k)}_{\mathFW}=A'^H_k.
\end{equation}

Using Eqs.~(\ref{A'k H}), (\ref{A'k N}) and (\ref{def}), the terms
$H_{FW}^{(k)}$ from $k=0$ to $k=12$ are given by
\begin{equation}
\begin{split}
&H'^{(0)}_{\mathFW}=0,~H'^{(1)}_{\mathFW}=-\left(\frac{g}{2}-1\right)\boldmu\cdot\mathbf{B},\\
&H'^{(2)}_{\mathFW}=-\left(\frac{g}{2}-1\right)\mex,\\
&H'^{(3)}_{\mathFW}=\frac{1}{2}\left(\frac{g}{2}-1\right)(\boldmu\cdot\boldxi)(\mathbf{B}\cdot\boldxi),\\
&H'^{(4)}_{\mathFW}=\left(\frac{g}{2}-1\right)\boldxi^2\mex,\\
&H'^{(5)}_{\mathFW}=-\frac{3}{8}\left(\frac{g}{2}-1\right)\boldxi^2(\boldmu\cdot\boldxi)(\mathbf{B}\cdot\boldxi),\\
&H'^{(6)}_{\mathFW}=-\frac{3}{8}\left(\frac{g}{2}-1\right)\boldxi^4\mex,\\
&H'^{(7)}_{\mathFW}=\frac{5}{16}\left(\frac{g}{2}-1\right)\boldxi^4(\boldmu\cdot\boldxi)(\mathbf{B}\cdot\boldxi),\\
&H'^{(8)}_{\mathFW}=\frac{5}{16}\left(\frac{g}{2}-1\right)\boldxi^6\mex,\\
&H'^{(9)}_{\mathFW}=-\frac{35}{128}\left(\frac{g}{2}-1\right)\boldxi^6(\boldmu\cdot\boldxi)(\mathbf{B}\cdot\boldxi),\\
&H'^{(10)}_{\mathFW}=-\frac{35}{128}\left(\frac{g}{2}-1\right)\boldxi^8\mex,\\
&H'^{(11)}_{\mathFW}=\frac{63}{256}\left(\frac{g}{2}-1\right)\boldxi^8(\boldmu\cdot\boldxi)(\mathbf{B}\cdot\boldxi),\\
&H'^{(12)}_{\mathFW}=\frac{63}{256}\left(\frac{g}{2}-1\right)\boldxi^{10}\mex.
\end{split}
\end{equation}
By using Eqs.~(\ref{Series1}) and (\ref{Series2}), the anomalous
Hamiltonian can be written as
\begin{equation}\label{HFW' total}
\begin{split}
H'_{\mathFW}&=\sum_{k=0}^{12}H'^{(k)}_{\mathFW}\\
&=-\left(\frac{g}{2}-1\right)\boldmu\cdot\mathbf{B}-\left(\frac{g}{2}-1\right)\frac{1}{\gammaop}\mex\\
&~~+\left(\frac{g}{2}-1\right)\left(1-\frac{\gammaop}{1+\gammaop}\right)\frac{1}{\gammaop}(\boldmu\cdot\boldxi)(\mathbf{B}\cdot\boldxi).
\end{split}
\end{equation}
Combining the FW transformed Dirac Hamiltonian [Eq.~(\ref{HFW
total})] and the anomalous Hamiltonain [Eq.~(\ref{HFW' total})],
we have (up to terms of $(\boldpi/mc)^{14}$)
\begin{equation}\label{HDP FW}
\begin{split}
\mathcal{H}_{\mathFW}&=H_{\mathFW}+H_{\mathFW}'\\
&=V+\gammaop
mc^2-\left(\frac{g}{2}-1+\frac{1}{\gammaop}\right)\boldmu\cdot\mathbf{B}\\
&~~+\left(\frac{g}{2}-\frac{\gammaop}{1+\gammaop}\right)\boldmu\cdot(\boost\times\mathbf{E})\\
&~~+\left(\frac{g}{2}-1\right)\frac{\gammaop}{1+\gammaop}(\boldmu\cdot\boost)(\mathbf{B}\cdot\boost).
\end{split}
\end{equation}
Equation (\ref{HDP FW}) is in agreement with the classical
Hamiltonian with $g\neq2$.

The FW transformed Dirac-Pauli Hamiltonian [Eq.~(\ref{HFW'
total})] can also be obtained by directly evaluating
Eq.~(\ref{HFWDP0}). Since Eq.~(\ref{HFWDP0}) is explicitly
hermitian, the calculation can be done without accounting for the
Dirac-Pauli string operator and the separation of hermitian and
anti-hermitian parts of the Dirac-Pauli energy operator
\cite{CLChang}. Up to $(\boldpi/mc)^{14}$, we find that the result
shown in Ref.~\cite{CLChang} is in agreement with the present
result. To find the classical correspondence of the quantum theory
of charged spin-1/2 particle, we have to perform the FW
transformation on the quantum Hamiltonian. The procedure presented
in this paper provides us a more systematic and efficient method
to obtain higher order expansion in the FW representation.

\section{Exact unitary transformation}\label{sec:EUT}

We now turn to the discussion of the \emph{exact} series expansions of the
Dirac and Dirac-Pauli generating operators. The exact unitary
transformation of a free particle Dirac Hamiltonian has been
given in Eq.~\ref{U-free}. In the presence
of electromagnetic fields, the series of successive FW
transformations becomes much more complicated. However, it is still possible to obtain the exact unitary transformation by deducing the close form from the finite-order series expansion, if the order we obtained is high enough. For example, the
exact unitary transformation of the free particle Dirac
Hamiltonian can be obtained from the successive FW transformations,
if terms in the series expansion is many enough to determine the
closed form. Therefore, in order to find the closed form for generic cases, we must proceed to higher orders. On
the other hand, it has been proposed that the low-energy and weak-field limit of the Dirac (resp. Dirac-Pauli) Hamiltonian is consistent
with the classical Hamiltonian, which is the sum of the classical
relativistic Hamiltonian and T-BMT Hamiltonian with $g=2$ (resp. $g\neq2$).
This suggests that there exists an exact unitary transformation for
the low-energy and weak-field limit. In this section we will find the closed
form of the unitary transformation from the high-order series
expansions of the generating operators.

The unitary transformation matrix is related to the
generating operator by Eqs. (\ref{U}) and (\ref{Def:YandZ}) in
Kutzelnigg's diagonalization method. If the closed form of
generating operator is found, the exact unitary transformation
matrix can be obtained.

For the low-energy and weak-field limit of the Dirac
Hamiltonian, the Dirac generating operator can be written as
\begin{equation}\label{EUT-X}
\begin{split}
X=\frac{X_1}{c}+\frac{X_3}{c^3}+\frac{X_5}{c^5}+\cdots.
\end{split}
\end{equation}
In Sec.~\ref{sec:HFW}, we have obtained the terms $X_{\ell}$ up to
order of $\ell=13$, which are given in Eq. (\ref{App:SolveX}). We
find that Eq. (\ref{EUT-X}) with Eq.
(\ref{App:SolveX}) can be incorporated into the closed form
\begin{widetext}
\begin{equation}\label{EUT-X-exact}
\begin{split}
X&=\frac{1}{1+\sqrt{1+(\boldsigma\cdot\boldxi)^2}}\boldsigma\cdot\boldxi+\left(\frac{1}{\sqrt{1+\boldxi^2}}-\frac{1}{1+\sqrt{1+\boldxi^2}}\right)\frac{-i}{mc^2}\boldmu\cdot\mathbf{E}+\left(\frac{1}{\sqrt{1+\boldxi^2}}\frac{1}{1+\sqrt{1+\boldxi^2}}\right)^2(\boldmu\cdot\boldxi)(\mathbf{E}\cdot\boldxi),
\end{split}
\end{equation}
\end{widetext}
The magnetic field generated from the operator
$(\boldsigma\cdot\boldpi)^2$ is included in the first term of Eq.
(\ref{EUT-X-exact}). In the absence of electromagnetic fields, the
kinetic momentum $\boldpi$ is replaced by the canonical momentum
$\mathbf{p}$. In this case, Eq. (\ref{EUT-X-exact}) becomes
$c\boldsigma\cdot\mathbf{p}/[mc^2+\sqrt{m^2c^4+c^2\mathbf{p}^2}]$,
which is the same as Eq. (\ref{free-X}), {and} the resulting unitary
transformation is exactly  Eq. (\ref{U-free}). We also note that in the absence of an electric
field, Eq. (\ref{EUT-X-exact}) becomes Eq. (\ref{Magnetic-X}).

Taking the anomalous magnetic moment into account, the
Dirac-Pauli generating operator can be written as
\begin{equation}
\mathcal{X}=X+X',
\end{equation}
where $X$ is given in Eq. (\ref{EUT-X-exact}). We find that the
anomalous generating operator $X'$ with Eq.
(\ref{X'k}) can be incorporated into the closed form
\begin{widetext}
\begin{equation}\label{EUT-X'-exact}
\begin{split}
X'&=\frac{X_3'}{c^3}+\frac{X_4'}{c^4}+\frac{X_5'}{c^5}+\cdots\\
&=\left(\frac{1}{\sqrt{1+\boldxi^2}}-\frac{1}{1+\sqrt{1+\boldxi^2}}\right)\left(\frac{g}{2}-1\right)\frac{1}{mc^2}\left(-i\boldmu\cdot\mathbf{E}+\frac{q\hbar}{2mc}\mathbf{B}\cdot\boldxi\right)\\
&~~+\frac{1}{\sqrt{1+\boldxi^2}}\left(\frac{1}{1+\sqrt{1+\boldxi^2}}\right)^2\left[-\frac{i}{mc^2}\left(\frac{g}{2}-1\right)(\boldmu\cdot\boldxi)(\mathbf{E}\cdot\boldxi)\right].
\end{split}
\end{equation}
\end{widetext}

The closed forms of $X$ and $X'$ have been deduced from the
high-order series expansions Eq. (\ref{App:SolveX}) and Eq.
(\ref{X'k}) respectively, but the rigorous proofs are stilling
missing. The merit of obtaining the closed forms is nevertheless
enormous: it allows us to guess the generic forms of $X_\ell$ and
$X'_\ell$ in the series expansions, which in turn enable us to
conduct rigorous proofs by mathematical induction
\cite{DWChiou2014}.

With Eqs. (\ref{EUT-X-exact}) and (\ref{EUT-X'-exact}) at hand, we
can formally construct the exact unitary transformation. However,
the main problem to be addressed is that the resulting exact
unitary matrix is valid only in the low-energy and weak-field
limit. In this regard, when we apply the exact unitary
transformation to the Dirac or Dirac-Pauli Hamiltonian, we have to
neglect nonlinear electromagnetic effects. In strong fields, the
particle's energy interacting with electromagnetic fields could
exceed the Dirac energy gap ($2mc^2$) and it is no longer adequate
to describe the relativistic quantum dynamics without taking into
account the field-theory interaction to the antiparticle. In fact,
some doubts have been thrown on the mathematical rigour of the FW
transformation \cite{Thaller1992}. The study of this paper
nevertheless suggests that the exact FW transformation indeed
exists and is valid in the low-energy and weak-field limit and
furthermore the FW transformed Hamiltonian agrees with the
classical counterpart (see \cite{DWChiou2014} for closer
investigations).

\section{Conclusions and Discussion}\label{sec:conclusions}

The motion of a particle endowed with charge and intrinsic spin is
governed by the classical Lorentz equation and the T-BMT
equation. Assuming that the canonical relation of classical spins
(via Poisson brackets) is the same as that of quantum spins (via
commutators), the T-BMT equation can be recast as the Hamilton's
equation and the T-BMT Hamiltonian is obtained. By treating
positions, momenta and spins as independent variables in pase
space, the classical Hamiltonian describing the motion of spin-1/2
charged particle is the sum of the classical relativistic
Hamiltonian and T-BMT Hamiltonian.

On the other hand, the correspondence between the classical
Hamiltonian and the low-energy and weak-field limit of Dirac equation
has been investigated by several authors. For a free particle, the
Foldy-Wouthuysen transformation of Dirac equation was shown to
exactly lead to the classical relativistic Hamiltonian of a free
particle. Intriguingly, when spin precession and interaction with
electromagnetic fields are also taken into account, it was found that
the connection between Dirac equation and classical Hamiltonian
becomes explicit if the order-by-order block diagonalization of
the the Dirac Hamiltonian can be proceed to higher-order terms.

The low-energy and weak-field limit of the relativistic quantum
theory of spin-1/2 charged particle is investigated by performing
the Kutzelnigg diagonalisation method on the Dirac Hamiltonian. We
show that in the presence of inhomogeneous electromagnetic fields
the Foldy-Wouthuysen transformed Dirac Hamiltonian up to terms
with $(\boldpi/mc)^4$ can be reproduced by the Kutzelnigg
diagonalisation method.

When the electromagnetic fields are homogeneous and nonlinear
effects are neglected, the Foldy-Wouthuysen transformation of the
Dirac Hamiltonian is obtained up to terms of $(\boldpi/mc)^{14}$.
The series expansion of the orbital part of the transformed Dirac
Hamiltonian in terms of the kinetic momentum enables us to define
the boost velocity operator. According to the correspondence
between the kinetic momentum and the boost velocity operator, we
found that up to terms of $(\boldpi/mc)^{14}$ the Foldy-Wouthuysen
transformed Dirac Hamiltonian is consistent with the classical
Hamiltonian with the gyromagnetic ratio given by $g=2$. Furthermore,
when the anomalous magnetic moment is considered as well, we found that up
to terms of $(\boldpi/mc)^{14}$ the Foldy-Wouthuysen transformed
Dirac-Pauli Hamiltonian is in agreement with the classical Hamiltonian
with $g\neq2$.

The investigation in this paper reveals the fact that the
classical Hamiltonian (classical relativistic Hamiltonian plus the
T-BMT Hamiltonian) must be the low-energy and weak-field limit of
the Dirac-Pauli equation. As shown in the above sections, we can
establish the connection order-by-order in the FW representation.
Moreover, this implies that, in the low-energy and weak-field
limit, there must exist an exact FW transformation that can
block-diagonalize the Dirac-Pauli Hamiltonian to the form
corresponding to the classical Hamiltonian. For a free particle,
the exact unitary transformation has been obtained by Foldy and
Wouthuysen, which alternatively can also be obtained by the
order-by-order method. We found that the generating operators can
be written as closed forms, and consequently we can formally
construct the exact unitary transformation that block-diagonalizes
the Dirac and Dirac-Pauli Hamiltonians. However, it should be
emphasized that the exact unitary transformation is valid only in
the low-energy and weak-field limit and existence of the exact
unitary transformation demands a rigours proof \cite{DWChiou2014}.

On the other hand, it is true that even if the unitary FW
transformation exists, it is far from unique, as one can easily
perform further unitary transformations which preserve the block
decomposition upon the block-diagonalized Hamiltonian (see also
\secref{sec:method}). While different block-diagonalization
transformations are unitarily equivalent to one another and thus
yield the same physics, however, the pertinent operators
$\boldsymbol{\sigma}$, $\mathbf{x}$, and $\mathbf{p}$ may
represent very different physical quantities in different
representations. To figure out the operators' physical
interpretations, it is crucial to compare the resulting FW
transformed Hamiltonian to the classical counterpart in a certain
classical limit via the \emph{correspondence principle}. In
Kutzelnigg's method, $\boldsymbol{\sigma}$ , $\mathbf{x}$, and
$\mathbf{p}$ simply represent the spin, position, and conjugate
momentum of the particle (as decoupled from the antiparticle) in
the resulting FW representation. In other words, Kutzelnigg's
method does not give rise to further transformations that obscure
the operators' interpretations other than block diagonalization.

The correspondence we observed may be extended to the case of
inhomogeneous electromagnetic fields (except that the Darwin term
has no classical correspondence) \cite{TWChen2013}, but
inhomogeneity gives rise to complications which make it cumbersome
to obtain the FW transformation in an order-by-order scenario,
including the Kutzelnigg method. We wish to tackle this problem in
further research.

\begin{acknowledgments}
The authors are grateful to C.-L.\ Chang for sharing his
calculations. T.W.C.\ would like to thank G.\ Y.\ Guo, R.\ Winkler and
M.-C.\ Chang for valuable discussions. T.W.C.\ is supported by the
National Science Council of Taiwan under Contract No.\ NSC
101-2112-M-110-013-MY3; D.W.C.\ is supported by the Center for
Advanced Study in Theoretical Sciences at National Taiwan
University.
\end{acknowledgments}

\appendix

\section{Hermiticity of FW transformed Dirac Hamiltonian}\label{App:Ham}

Under the unitary transformation [Eq.~(\ref{UHU})], the
Foldy-Wouthuysen transformed Dirac Hamiltonian is given by the
upper-left term of $UH_DU^{\dag}$, which is
\begin{equation}\label{App:HFW}
\begin{split}
H_{\mathFW}&=\left(Yh_++YX^{\dag}h_0\right)Y+\left(Yh_0+YX^{\dag}h_-\right)XY\\
&=Y\left(h_++X^{\dag}h_0+h_0X+X^{\dag}h_-X\right)Y.
\end{split}
\end{equation}
Since the operators $Y$, $h_+$ and $h_0$ are hermitian, it is easy
to show that $H_{FW}$ also satisfies $H_{FW}=H_{FW}^{\dag}$. The
two off-diagonal terms are given by
\begin{equation}\label{App:HFWX}
\begin{split}
&H_{X}=Z\left(-Xh_++h_0-Xh_0X+h_-X\right)Y,\\
&H_{X^{\dag}}=Y\left(-h_+-X^{\dag}h_0X^{\dag}+h_0+X^{\dag}h_-\right)Z.
\end{split}
\end{equation}
Equation (\ref{App:HFW}) can be further simplified by using
$H_{X}=0$ and $H_{X^{\dag}}=0$.

The brackets in the second equality of Eq.~(\ref{App:HFW}) can be
rewritten as
\begin{equation}\label{App:HFW1}
\begin{split}
&\left(h_++X^{\dag}h_0+h_0X+X^{\dag}h_-X\right)\\
&=V+mc^2+X^{\dag}h_0+h_0X+X^{\dag}\left(V-mc^2\right)X\\
&=V+h_0X+mc^2\left(1-X^{\dag}X\right)+\left(X^{\dag}VX+X^{\dag}h_0\right).
\end{split}
\end{equation}
On the other hand, we have
\begin{equation}\label{App:HFW2}
\begin{split}
&\left(X^{\dag}VX+X^{\dag}h_0\right)\\
&=X^{\dag}\left(VX+h_0\right)\\
&=X^{\dag}\left([V,X]+XV+h_0\right)\\
&=X^{\dag}\left(2mc^2X+Xh_0X+XV\right)\\
&=2mc^2X^{\dag}X+X^{\dag}Xh_0X+X^{\dag}XV,
\end{split}
\end{equation}
where Eq.~(\ref{EqX}) was used in the third equality of
Eq.~(\ref{App:HFW2}). Substituting Eq.~(\ref{App:HFW2}) into
Eq.~(\ref{App:HFW1}), we have
\begin{equation}\label{App:HFW3}
\begin{split}
&\left(h_++X^{\dag}h_0+h_0X+X^{\dag}h_-X\right)\\
&=V+h_0X+mc^2\left(1-X^{\dag}X\right)+2mc^2X^{\dag}X\\
&~~+X^{\dag}Xh_0X+X^{\dag}XV\\
&=\left(1+X^{\dag}X\right)V+\left(1+X^{\dag}X\right)h_0X+mc^2\left(1+X^{\dag}X\right)\\
&=Y^{-2}\left(V+h_0X+mc^2\right).
\end{split}
\end{equation}
Inserting Eq.~(\ref{App:HFW3}) into Eq.~(\ref{App:HFW}), we obtain
\begin{equation}\label{App:HFW4}
\begin{split}
H_{\mathFW}&=YY^{-2}\left(V+h_0X+mc^2\right)Y\\
&=mc^2+Y^{-1}\left(V+h_0X\right)Y.
\end{split}
\end{equation}
The condition $H_{X^{\dag}}=0$ implies
\begin{equation}\label{App:EqXdag}
X^{\dag}=\frac{1}{2mc^2}\left(h_0-X^{\dag}h_0X^{\dag}+[X^{\dag},V]\right).
\end{equation}
Applying Eq.~(\ref{App:EqXdag}) to Eq.~(\ref{App:HFW}), we have
\begin{equation}
\begin{split}
H_{\mathFW}&=Y\left(h_++X^{\dag}h_0+h_0X+X^{\dag}h_-X\right)Y\\
&=Y\left[V+X^{\dag}h_0+mc^2(1-X^{\dag}X)+\left(X^{\dag}V+h_0\right)X\right]Y\\
&=Y[V+X^{\dag}h_0+mc^2(1-X^{\dag}X)+2mc^2X^{\dag}X\\
&~~+VX^{\dag}X+X^{\dag}h_0X^{\dag}X]Y\\
&=Y\left(VY^{-2}+X^{\dag}h_0Y^{-2}+mc^2Y^{-2}\right)Y,
\end{split}
\end{equation}
where Eq.~(\ref{App:EqXdag}) was used in the second equality. We
obtain
\begin{equation}\label{App:HFW5}
H_{\mathFW}=mc^2+Y\left(V+X^{\dag}h_0\right)Y^{-1}.
\end{equation}
Because $Y$, $h_0$ and $V$ are hermitian operators, this implies
that the hermitian of Eq.~(\ref{App:HFW4}) is
$H^{\dag}_{\mathFW}=mc^2+Y\left(V+X^{\dag}h_0\right)Y^{-1}$, and
this is the same as Eq.~(\ref{App:HFW5}). As a consequence, we
have $H_{\mathFW}^{\dag}=H_{\mathFW}$.

\section{Hermiticity of FW transformed Dirac-Pauli Hamiltonian}\label{App:Ham2}
In this appendix, we will show that the FW transformed Dirac-Pauli
Hamiltonian can be written as Eq.~(\ref{H FWDP}) and show that
Eq.~(\ref{H FWDP}) is a hermitian operator. Under the unitary
transformation [Eq.~(\ref{UHDPU})], the Foldy-Wouthuysen
transformed Dirac-Pauli Hamiltonian is given by the upper-left
term of $U\mathcal{H}U^{\dag}$:
\begin{equation}\label{App:HDP FW}
\mathcal{H}_{\mathFW}=\mathcal{Y}\left(H_++\mathcal{X}^{\dag}H^{\dag}_0+H_0\mathcal{X}+\mathcal{X}^{\dag}H_-\mathcal{X}\right)\mathcal{Y},
\end{equation}
where $H_+=V+V_B+mc^2$, $H_0=h_0+iV_E$ and $H_=V-V_B-mc^2$. The
operator $h_0$ is $h_0=c\,\sdotp$. Since the operator
$\mathcal{Y}$ is hermitian, it is easy to show that
Eq.~(\ref{App:HDP FW}) also satisfies
$\mathcal{H}_{\mathFW}=\mathcal{H}_{\mathFW}^{\dag}$. The two
off-diagonal terms are required to vanish and they are given by
\begin{equation}\label{App:HDP FWX1}
-\mathcal{X}H_++H_0^{\dag}-\mathcal{X}H_0\mathcal{X}+H_-\mathcal{X}=0,
\end{equation}
and
\begin{equation}\label{App:HDP FWX2}
-H_+\mathcal{X}^{\dag}-\mathcal{X}^{\dag}H_0^{\dag}\mathcal{X}^{\dag}+H_0+\mathcal{X}^{\dag}H_-=0.
\end{equation}
By multiplying $\mathcal{X}^{\dag}$ on the left-hand side of Eq.~(\ref{App:HDP FWX1}), we have
\begin{equation}\label{App:HDP FWX3}
\left(\mathcal{X}^{\dag}H_0^{\dag}+\mathcal{X}^{\dag}H_-\mathcal{X}\right)=\mathcal{X}^{\dag}\mathcal{X}H_++\mathcal{X}^{\dag}\mathcal{X}H_0\mathcal{X}.
\end{equation}
Substituting Eq.~(\ref{App:HDP FWX3}) into Eq.~(\ref{App:HDP FW})
by eliminating
$\left(\mathcal{X}^{\dag}H_0^{\dag}+\mathcal{X}^{\dag}H_-\mathcal{X}\right)$,
we obtain
\begin{equation}\label{App:HDP FW1}
\mathcal{H}_{\mathFW}=\mathcal{Y}^{-1}\left(H_++H_0\mathcal{X}\right)\mathcal{Y},
\end{equation}
where the definition of the operator
$\mathcal{Y}=1/\sqrt{1+\mathcal{X}^{\dag}{X}}$ was used. On the
other hand, multiplying $\mathcal{X}$ on the right-hand side of
Eq.~(\ref{App:HDP FWX2}), we have
\begin{equation}\label{App:HDP FWX4}
\left(H_0\mathcal{X}+\mathcal{X}^{\dag}H_-\mathcal{X}\right)=H_+\mathcal{X}^{\dag}\mathcal{X}+\mathcal{X}^{\dag}H_0^{\dag}\mathcal{X}^{\dag}\mathcal{X}.
\end{equation}
Substituting Eq.~(\ref{App:HDP FWX4}) into Eq.~(\ref{App:HDP FW})
and eliminating the term
$\left(H_0\mathcal{X}+\mathcal{X}^{\dag}H_-\mathcal{X}\right)$, we
obtain
\begin{equation}\label{App:HDP FW2}
\mathcal{H}_{\mathFW}=\mathcal{Y}\left(H_++\mathcal{X}^{\dag}H_0^{\dag}\right)\mathcal{Y}^{-1}.
\end{equation}
Because $\mathcal{Y}$ is a hermitian operator and so is $H_+$,
this implies that the hermitian of Eq.~(\ref{App:HDP FW1}) is
$\mathcal{H}^{\dag}_{\mathFW}=\mathcal{Y}\left(H_0+\mathcal{X}^{\dag}H_0^{\dag}\right)\mathcal{Y}^{-1}$,
which is the same as Eq.~(\ref{App:HFW5}). As a consequence, we
have $\mathcal{H}_{\mathFW}^{\dag}=\mathcal{H}_{\mathFW}$. On the
other hand, $H_++H_0\mathcal{X}$ can be written as
$\left(H_++H_0\mathcal{X}\right)=mc^2+V+V_B+(h_0+iV_E)\mathcal{X}$.
Equation (\ref{App:HDP FW1}) can be simplified as
\begin{equation}
\mathcal{H}_{\mathFW}=mc^2+e^{\mathcal{G}/2}\mathcal{A}e^{-\mathcal{G}/2},
\end{equation}
where the operators $\mathcal{A}$ and $\mathcal{G}$ are defined as
$\mathcal{A}=V+h_0\mathcal{X}+V_B+iV_E\mathcal{X}$ and
$\mathcal{G}=\ln\left(1+\mathcal{X}^{\dag}\mathcal{X}\right)$,
respectively.


\end{document}